\documentclass[twocolumn]{aastex63} 

\usepackage[caption = false]{subfig}
\usepackage{graphicx}

\usepackage{xspace}

\newcommand\hcop{HCO$^+$\xspace}
\newcommand\htcop{H$^{13}$CO$^+$\xspace}

\newcommand\kms{km s$^{-1}$\xspace}

\begin{document}

\title{The young embedded disk L1527 IRS: constraints on the water snowline and cosmic ray ionization rate from HCO$^+$ observations}

\correspondingauthor{Merel L.R. van 't Hoff}
\email{mervth@umich.edu}

\author{Merel L.R. van 't Hoff}
\affil{Department of Astronomy, University of Michigan, 1085 S. University Ave., Ann Arbor, MI 48109-1107, USA}

\author{Margot Leemker}
\affil{Leiden Observatory, Leiden University, P.O. Box 9513, 2300 RA Leiden, The Netherlands}

\author{John J. Tobin}
\affil{National Radio Astronomy Observatory, 520 Edgemont Rd., Charlottesville, VA 22903, USA}

\author{Daniel Harsono} 
\affil{Leiden Observatory, Leiden University, P.O. Box 9513, 2300 RA Leiden, The Netherlands}
\affil{Institute of Astronomy and Astrophysics, Academia Sinica, P.O. Box 23-141, Taipei 106, Taiwan} 

\author{Jes K. J{\o}rgensen}
\affil{Niels Bohr Institute, University of Copenhagen, {\O}ster Voldgade 5-7, 1350 Copenhagen K., Denmark}

\author{Edwin A. Bergin}
\affil{Department of Astronomy, University of Michigan, 1085 S. University Ave., Ann Arbor, MI 48109-1107, USA}




\begin{abstract}

\noindent The water snowline in circumstellar disks is a crucial component in planet formation, but direct observational constraints on its location remain sparse due to the difficulty of observing water in both young embedded and mature protoplanetary disks. Chemical imaging provides an alternative route to locate the snowline, and \hcop isotopologues have been shown to be good tracers in protostellar envelopes and Herbig disks. Here we present $\sim$0$\farcs$5 resolution ($\sim$35 au radius) Atacama Large Millimeter/submillimeter Array (ALMA) observations of \hcop $J=4-3$ and \htcop $J=3-2$ toward the young (Class 0/I) disk L1527 IRS. Using a source-specific physical model with the midplane snowline at 3.4 au and a small chemical network, we are able to reproduce the \hcop and \htcop emission, but for \hcop only when the cosmic ray ionization rate is lowered to $10^{-18}$ s$^{-1}$. Even though the observations are not sensitive to the expected \hcop abundance drop across the snowline, the reduction in \hcop above the snow surface and the global temperature structure allow us to constrain a snowline location between 1.8 and 4.1 au. Deep observations are required to eliminate the envelope contribution to the emission and to derive more stringent constraints on the snowline location. Locating the snowline in young disks directly with observations of H$_2$O isotopologues may therefore still be an alternative option. With a direct snowline measurement, \hcop will be able to provide constraints on the ionization rate. 

\end{abstract}




\section{Introduction} \label{sec:intro}

Evidence for an early start of planet formation, when the disk is still embedded in its envelope, has been accumulating. For example, rings in continuum emission that are ubiquitously observed toward Class II protoplanetary disks \citep[e.g.,][]{Andrews2018} and could be a signpost of forming planets \citep[e.g.,][]{Bryden1999,Zhu2014,Dong2018}, are now also observed in disks as young as only $\sim$0.5 Myr \citep{ALMAPartnership2015,Segura-Cox2020,Sheehan2020}. Evidence for grain growth beyond interstellar medium (ISM) sizes has been inferred from low dust opacity spectral indexes in Class 0 sources \citep{Kwon2009,Shirley2011}, dust polarization \citep[e.g.,][]{Kataoka2015,Kataoka2016,Yang2016}, decreasing dust masses derived from (sub-)millimeter observations for more evolved systems \citep[e.g.][]{Tychoniec2020}, and CO isotopologue emission \citep{Harsono2018}. In addition, outflows present in this early phase may provide a way to overcome the radial drift barrier \citep{Tsukamoto2021}. 

One of the key parameters in planet-formation models is the location of the water snowline, that is, the disk midplane radius at which water molecules freeze out onto the dust grains. At this location, the growth of dust grains, and thus the planet formation efficiency, is expected to be significantly enhanced through triggering of the streaming instability \citep[e.g.,][]{Stevenson1988,Schoonenberg2017,Drazkowska2017}. In addition, since water is the dominant carrier of oxygen, the elemental carbon-to-oxygen (C/O) ratio of the planet forming material changes across the water snowline \citep{Oberg2011,Eistrup2018}.  \citet{Lichtenberg2021} illustrated the importance of the snowline location during disk evolution as migration of the snowline may be an explanation for the isotopic dichotomy of Solar System meteorites \citep[e.g.,][]{Leya2008,Trinquier2009,Kruijer2017}. In a different perspective, theoretical studies have shown that the position of the water snowline depends on the disk viscosity and dust opacity \citep{Davis2005,Lecar2006,Garaud2007,Oka2011}, hence snowline measurements will provide important information for disk evolution models. Overall, observational constraints on the snowline location are thus crucial to understand planet formation and its outcome, and observations of young disks are particularly important as they represent the earliest stages in planet formation.

Unfortunately, water emission is difficult to detect in both young and more evolved disks \citep{Du2017,Notsu2018,Notsu2019,Harsono2020}, and thus determining the exact location of the snowline is challenging. However, observations of protostellar envelopes have shown that \htcop can be used as an, indirect, chemical tracer of the water snowline \citep{Jorgensen2013,vantHoff2018a,vantHoff2022,Hsieh2019}. This is based on gaseous water being the most abundant destroyer of \hcop in warm dense gas around young stars. \hcop is therefore expected to be abundant only in the region where water is frozen out and gaseous CO is available for its formation \citep{Phillips1992,Bergin1998}. The high optical depth of the main isotopologue, \hcop, impedes snowline measurements in protostellar envelopes \citep{vantHoff2022}, warranting the use of the less abundant isotopologues \htcop or HC$^{18}$O$^+$. Modeling of \hcop emission from Herbig disks has shown that this optical depth problem is partly mitigated in disks due to their Keplerian velocity pattern, as different velocities trace different radii \citep{Leemker2021}. 

Here, we present Atacama Large Millimeter/submillimeter Array (ALMA) observations of \hcop and \htcop in the young disk L1527 IRS (also known as IRAS 04368+2557 and hereafter referred to as L1527). This well-studied Class 0/I protostar located in the Taurus molecular cloud (142 pc, \citealt{GAIAcollaboration2021,Krolikowski2021}) is surrounded by a 75--125 au Keplerian disk \citep{Tobin2012,Tobin2013,Aso2017} that is viewed nearly edge-on \citep{Tobin2008,Oya2015} and is embedded in an extended envelope \citep[e.g.,][]{Ohashi1997,Tobin2008}. The observations are described and presented in Sects.~\ref{sec:Observations} and \ref{sec:Results}, respectively. In Sect.~\ref{sec:Modeling} we use the physical structure for L1527 derived by \citet{Tobin2013} to model the \hcop abundance, and \hcop and \htcop emission, incorporating the \hcop abundance through either simple parametrization (Sect.~\ref{sec:ParametrizedModel}) or the use of a small chemical network (Sect.~\ref{sec:ChemicalModel}). In Sect.~\ref{sec:SnowlineLocation} we then use the chemical modeling results to constrain the water snowline location in L1527. Finally, we discuss the cosmic ray (CR) ionization rate in Sect.~\ref{sec:CRrate} and summarize the main conclusions in Sect.~\ref{sec:Conclusions}.


\section{Observations} \label{sec:Observations} 

L1527 was observed with ALMA in \hcop on 2014 June 14 (project code 2012.1.00346.S, PI: N. Evans) for a total on source time of 11 minutes. These observations were carried out using 33 antennas sampling baselines up to 650 m.  The correlator setup consisted of four 234 MHz spectral windows, including one targeting the \hcop $J=4-3$ transition at 356.734223 GHz, with 61 kHz ($\sim$0.05 km~s$^{-1}$) spectral resolution. 

In addition, L1527 was observed in \htcop on 2015 August 11 and 12 and September 2 (project code 2012.1.00193.S, PI: J.J. Tobin) for a total of 43 minutes on source per execution ($\sim$2.2 hours total). The observations were carried out with 42, 44 and 34 antennas for the three respective observing dates and sampled baselines up to 1.6 km. The correlator setup contained two 117 MHz spectral windows, including one targeting the \htcop $J=3-2$ transition at 260.255339 GHz, with 31 kHz ($\sim$0.05 km~s$^{-1}$) spectral resolution and two 2 GHz spectral windows with 15.6 MHz resolution, aimed for continuum measurements. 

Calibration, self-calibration and imaging of the \hcop and \htcop datasets were done using versions 4.2.1 and 4.3.1 of the Common Astronomy Software Application (CASA, \citealt{McMullin2007}), respectively, where the \hcop data were calibrated using the ALMA Pipeline. For the \hcop observations, J0510+1800 was used as bandpass, phase and flux calibrator. For the \htcop observations, the bandpass calibrator was J0423--0120, the flux calibrator was J0423--0130, and the phase calibrator was J0510+1800 for the August observations and J0440+2728 for the September observations. Both lines are imaged at a spectral resolution of 0.1 \kms. A uv taper of 500 k$\lambda$ was applied to increase the signal-to-noise ratio of the \htcop image cube. The restoring beam is 0$\farcs$50 $\times$ 0$\farcs$30 (PA = -3.2$^\circ$) for \hcop and 0$\farcs$47 $\times$ 0$\farcs$28 (44.7$^\circ$) for H$^{13}$CO$^+$, and the images have an rms of 20 mJy beam$^{-1}$ channel$^{-1}$ and 3.9 mJy beam$^{-1}$ channel$^{-1}$, respectively.  The maximum recoverable scale is 2$\farcs$7 (380 au) for the \hcop observations and 2$\farcs$0 (280 au) for \htcop, that is, spanning the disk (75--125 au; \citealt{Tobin2012,Tobin2013,Aso2017}) and innermost envelope. 

\begin{figure}
\centering
\includegraphics[trim={0cm 5.2cm 4cm 1.4cm},clip]{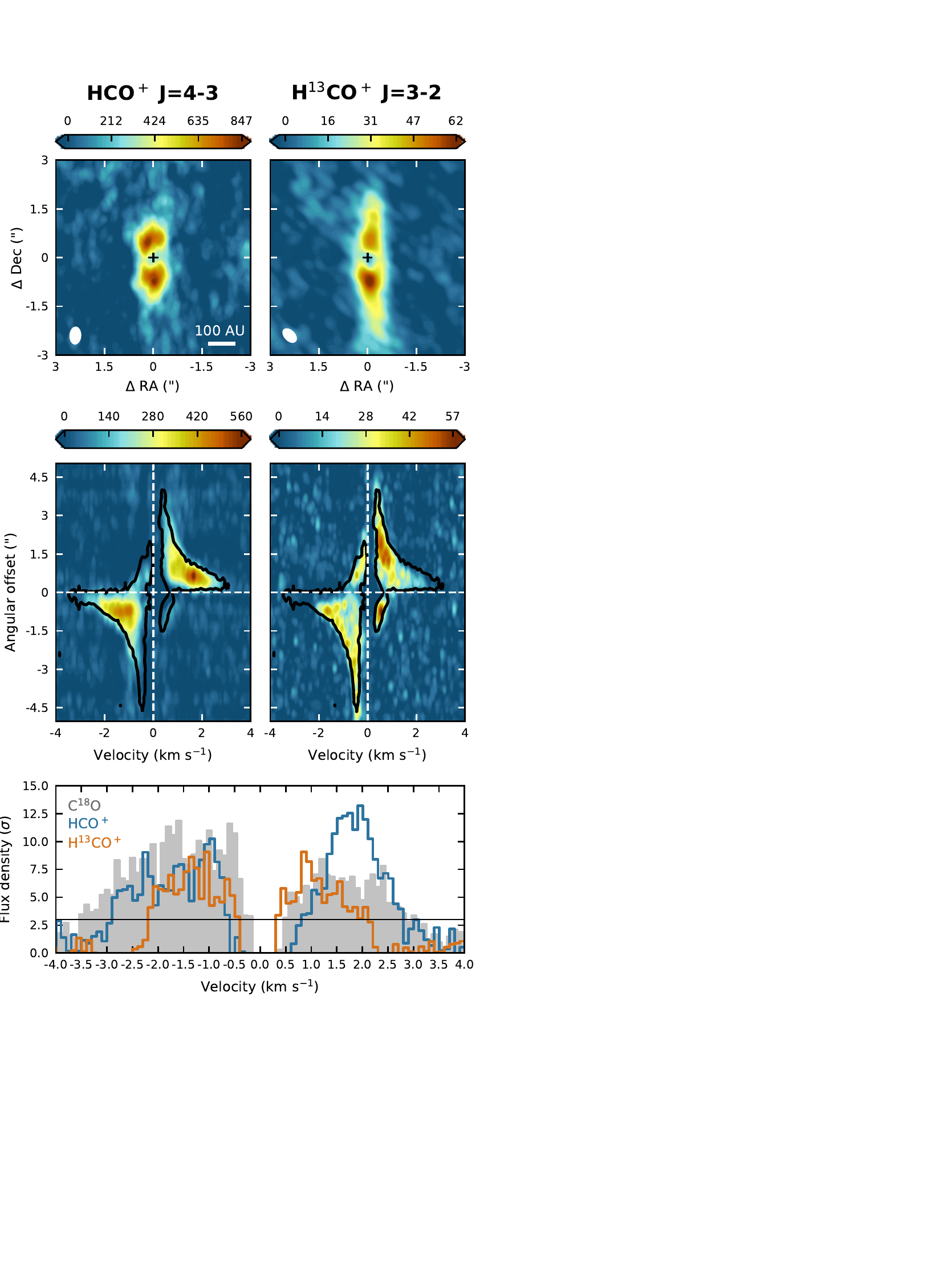}
\caption{Integrated intensity maps (top) and position-velocity (\textit{pv}) diagrams (middle) for the \hcop $J=4-3$ (left) and \htcop $J=3-2$ (right) transitions toward L1527. Central velocity channels ($\Delta v \leq |0.5|$ \kms) with resolved out emission are omitted from the integrated intensity maps. The velocity axis of the \textit{pv} diagrams is centered on the systemic velocity of 5.9 \kms and the C$^{18}$O $J=2-1$ \textit{pv} diagram is overlaid in black contours (3$\sigma$).  The color scale is in mJy beam$^{-1}$ km s$^{-1}$ for the integrated intensity maps and in mJy beam$^{-1}$ for the \textit{pv} diagrams. The beam is shown in the bottom left corner of the top panels and the velocity resolution is 0.1 km s$^{-1}$. The bottom panel shows cuts through the \textit{pv} diagrams close to the midplane ($0\farcs2$ and -$0\farcs2$ for, respectively, redshifted and blueshifted C$^{18}$O and \hcop emission, and $\pm 0\farcs5$ for \htcop) to highlight the difference in velocity extent between C$^{18}$O (solid grey), \hcop (blue line) and \htcop (orange line). The flux is expressed in factors of $\sigma$ for each dataset, and the horizontal line marks the $3\sigma$ level.  }
\label{fig:L1527_observations}
\end{figure}


\section{Results} \label{sec:Results}

Figure~\ref{fig:L1527_observations} (top panels) presents the integrated intensity maps for \hcop $J=4-3$ and \htcop $J=3-2$ toward L1527. Emission from channels near the systemic velocity ($\Delta v \leq |0.5|$ \kms) where most of the emission is resolved out are omitted. Both molecules display emission elongated along the north-south direction, that is, along the major axis of the edge-on disk, with the blueshifted emission south of the protostar. The \hcop emission is radially more compact than the \htcop emission, likely because the $J=4-3$ transition traces warmer and denser material than the $J=3-2$ transition. The higher sensitivity of the \htcop observations and more resolved out emission for the optically thicker \hcop emission possibly play a role as well. For both lines, a central depression is visible, which at first thought may be interpreted as a lack of \hcop and \htcop in the inner region of the disk. However, modeling of \hcop emission by \citet{Hsieh2019} showed that a ring-shaped distribution of \hcop molecules in an embedded disk does not result in a central depression in emission for highly inclined sources. For the edge-on disk L1527 the central depressions are thus due to a combination of optically thick continuum emission in the central beam, resolved out line emission and the subtraction of continuum from optially thick line emission. 

A better picture of the spatial origin of the emission can be obtained from position-velocity (\textit{pv}) diagrams as shown in Fig.~\ref{fig:L1527_observations} (middle panels). In principle, in these diagrams, disk emission is located at small angular offsets and high velocities, while envelope emission extends to larger offsets but has lower velocities. The \textit{pv} diagrams show that the \hcop emission peaks at angular offsets of $\sim$1$^{\prime\prime}$ and velocities between $\sim$1--2 km s$^{-1}$, while the \htcop emission peaks at larger offsets ($\sim$1.5--3$^{\prime\prime}$) and lower velocities ($\lesssim$1 km s$^{-1}$). The presence of an infalling envelope is also evident from the presence of redshifted emission on the predominantly blueshifted south side of the source and blueshifted emission in the north. These components are strongest for H$^{13}$CO$^+$. Together, this suggests that the \hcop emission is dominated by the disk and innermost envelope and that the \htcop emission originates mostly at larger radii ($\gtrsim$ 140 au). However, if the emission is optically thick, emission observed at small spatial offsets from source center may in fact originate at much larger radii (see e.g., \citealt{vantHoff2018b}), so the difference between \hcop and \htcop can be partially due to an optical depth effect. 

An absence of \hcop inside the water snowline in the inner disk would show up in the \textit{pv} diagram as an absence of emission at the highest velocities. Because at these highest velocities only emission from the disk, and not from the envelope, is present (see e.g., Fig.~\ref{fig:VelocityField}) this effect can still be visible even if the emission becomes optically thick in the envelope. As a reference, the 3$\sigma$ contour of C$^{18}$O $J=2-1$ emission at comparable resolution ($0\farcs43 \times 0\farcs28$) is overlaid on the \hcop and \htcop \textit{pv} diagrams. These C$^{18}$O observations were previously presented by \citet{vantHoff2018b}, but to maximize the signal-to-noise ratio, we show here the combined data from the long and short baseline tracks of the observing program, while \citet{vantHoff2018b} only used the long baseline executions. C$^{18}$O is present throughout the entire disk, so an absence of \hcop and \htcop emission at the highest C$^{18}$O velocities signals a depression or absence of these molecules in the inner region of the disk. The highest blue- and redshifted velocities observed for C$^{18}$O are $-3.6$ \kms and $+3.0$ \kms, respectively, with respect to the source velocity. \hcop reaches velocities close to the highest redshifted C$^{18}$O velocity, that is, $-2.8$ and $+2.9$ km s$^{-1}$, while \htcop is confined between $-2.1$ and $2.2$ km s$^{-1}$ at the 3$\sigma$ level of the observations (see Fig.\ref{fig:L1527_observations}, bottom panel). 

A more quantitative constraint on the spatial origin of the emission can be set by considering the velocity structure. To calculate the velocity field, we adopt a Keplerian rotating disk with an outer radius of 125 au \citep{Tobin2013} embedded in a rotating infalling envelope following the prescription by \citet{Ulrich1976} and \citet{Cassen1981}. We use a stellar mass of 0.4 $M_\odot$ as this was found to best reproduce ALMA observations of $^{13}$CO and C$^{18}$O \citep{vantHoff2018b}. This is slightly lower than the $\sim$0.45 $M_\odot$ derived by \citet{Aso2017}. The resulting midplane velocity field is displayed in Fig.~\ref{fig:VelocityField}. For this stellar mass and disk radius, emission at velocities $\gtrsim |2.6|$ km s$^{-1}$ offset from the source velocity originates solely in the disk. The highest velocity \hcop emission observed at the current sensitivity is predominantly coming from the disk at radii $\gtrsim$ 42 au. All H$^{13}$CO$^+$ velocity channels contain emission from both disk and envelope. This means that either the observed H$^{13}$CO$^+$ emission originates solely in the envelope, or there is some emission coming from the outer disk (radii $\gtrsim$ 73 au) as well. As illustrated in Fig.~\ref{fig:VelocityField}, these cases are not trivial to distinguish as the envelope velocity profile results in envelope emission being present at small angular offsets from the protostellar position. However, taken together, these results thus suggest an absence of \hcop emission in the inner $\sim$40 au at the sensitivity of our observations. 

\begin{figure*}
\centering
\includegraphics[trim={0cm 12.8cm 0cm 0cm},clip]{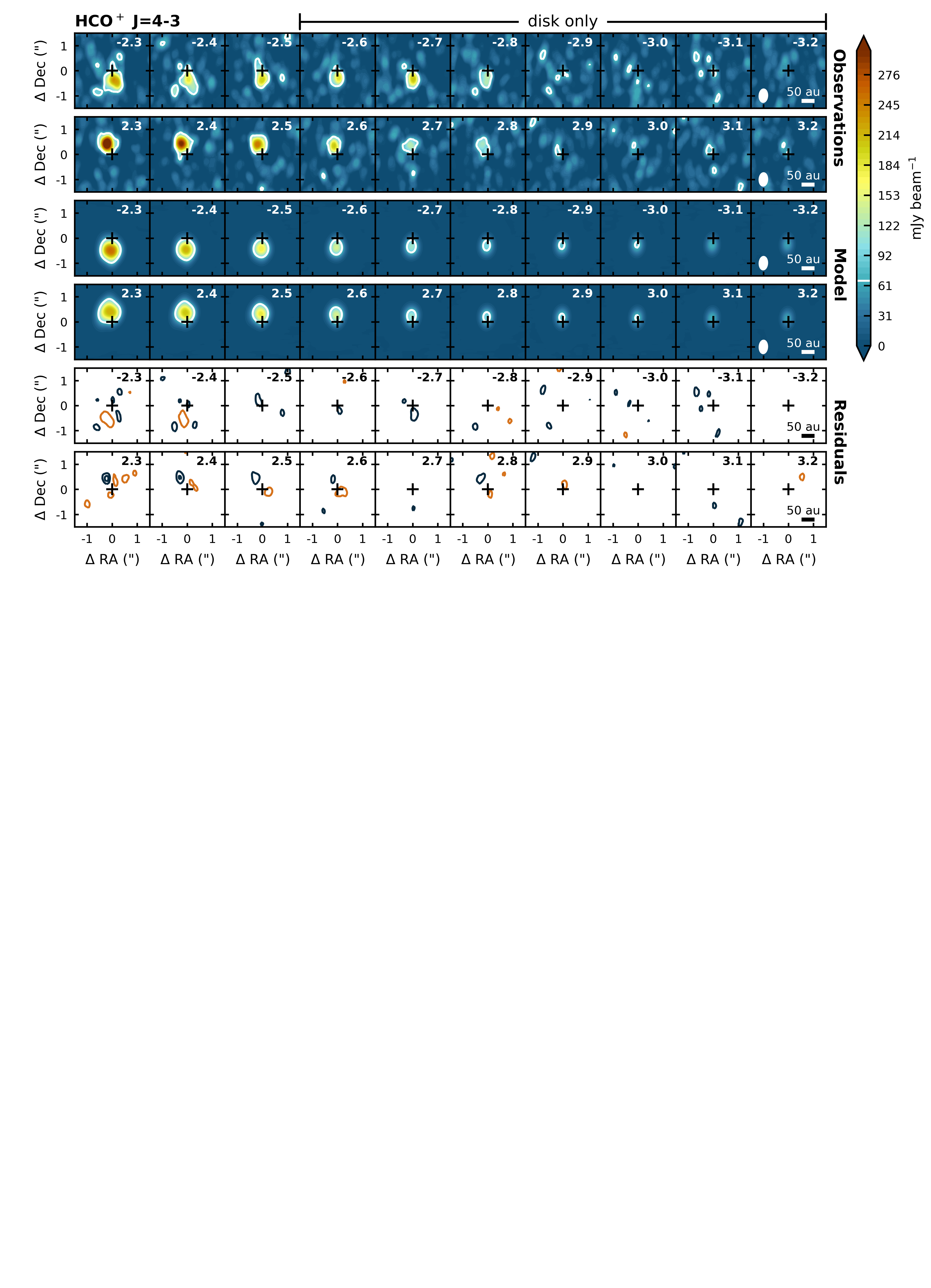}
\caption{Selected channels of \hcop $J=4-3$ emission from observations (top two rows) and from a L1527 specific model with an abundance of $2\times10^{-11}$ at radii $\leq$60 au and an abundance of $2\times10^{-10}$ at larger radii (middle two rows). The velocities offset from the source velocity (km s$^{-1}$) are listed in the top right corner of each panel, and channels at velocities $\gtrsim|2.6|$ km s$^{-1}$ contain only emission from the disk. A white contour denotes the 3$\sigma$ level. Residuals after subtracting the model from the observations are shown in the bottom two rows. Black contours are in steps of 3$\sigma$, starting at 3$\sigma$ and orange contours are in steps of -3$\sigma$ starting at -3$\sigma$. The black cross marks the continuum peak and the beam is shown in bottom left corner of the right most panels. }
\label{fig:L1527_HCO+model}
\end{figure*}


\section{Modeling of the HCO$^+$ emission}\label{sec:Modeling}

To further interpret these observations, we make synthetic \hcop and \htcop images using the physical structure for L1527 derived by \citet{Tobin2013} and that was also used by \citet{vantHoff2018b} and \citet{vantHoff2020} to model the $^{13}$CO, C$^{18}$O and C$^{17}$O emission. In short, this model contains a 125 au Keplerian disk within a rotating infalling envelope \citep{Ulrich1976,Cassen1981} and is the result of fitting a large grid of 3D radiative transfer models to the thermal dust emission in the (sub-)millimeter, the scattered light $L^{\prime}$ image, and the multi-wavelength SED. In order to fit the multi-wavelength continuum emission, a parameterized sub/millimeter dust opacity was adopted with a value of 3.5 cm$^{2}$ g$^{-1}$ at 850 $\mu$m \citep{Andrews2005} and the best fit model has a dust opacity spectral index $\beta$ of 0.25. This dust opacity suggests that some grain growth has occured (see \citealt{Tobin2013} for more discussion). In our model, the dust then becomes optically thick at radii $\lesssim$4 au for different angular offsets along the disk major axis at the frequency of the \hcop $J=4-3$ transition (356.734288 GHz) (see Fig.~\ref{fig:VelocityField}). The temperature and density structure of the model is shown in Fig.~\ref{fig:PhysicalStructure}.

We employ two approaches to constrain the spatial origin of the \hcop and \htcop emission and the water snowline location. First, we adopt a parametrized abundance structure where the \hcop abundance is vertically constant but can change at different radii (Sect.~\ref{sec:ParametrizedModel}). This simple type of model will allow us to address whether the non-detection of \hcop and \htcop emission at velocities as high as observed for C$^{18}$O is due to a steep drop in abundance, as expected inside the water snowline. Second, we use a small chemical network for \hcop as presented by \citet{Leemker2021} for a more detailed study of the snowline location (Sect~\ref{sec:ChemicalModel}). In both cases, image cubes are simulated with the 3D radiative transfer code LIME \citep{Brinch2010}, assuming LTE and using molecular data files from the LAMDA database \citep{Schoier2005,vanderTak2020}. The synthetic image cubes are continuum subtracted and convolved with the observed beam size.


\subsection{Parametrized abundance structure} \label{sec:ParametrizedModel}

Our goal here is to determine whether the absence of \hcop and \htcop emission in the inner disk is due to a sharp drop in abundance, as expected inside the water snowline. We therefore parametrize the \hcop abundance as function of radius and focus on the intermediate and high velocity channels that contain emission from the disk and inner envelope. 

Velocity-channel emission maps of a model that reproduces the \hcop emission at velocities $\geq |2.3|$ \kms reasonably well are presented in Fig.~\ref{fig:L1527_HCO+model}. This model has a \hcop abundance of $2\times10^{-11}$ at radii $\leq$ 60 au, and an abundance of $2\times10^{-10}$ at larger radii. This latter abundance is not high enough to reproduce the observed envelope emission at lower velocities, and this is most likely the reason that the redshifted emission at intermediate velocities (2.3--2.5 \kms) is slightly underestimated. However, the important result here is that the \hcop abundance inside 60 au is low, and therefore, the non-detection of emission at velocities $\geq |2.9|$ \kms (tracing the inner $\sim$40 au) could be due to the sensitivity of the observations. Abundances higher than $2\times10^{-11}$ produce emission above the observed 3$\sigma$ level at velocities $\geq |2.9|$ \kms, but a further drop in abundance at radii $\lesssim$40 au cannot be assessed. 

The abundance in the outer disk ($>$ 60 au) is hard to constrain as well, because the abundance in the outer disk and inner envelope are degenerate. A model with an abundance of $2\times10^{-11}$ throughout the entire disk and an envelope abundance of $1\times10^{-9}$ reproduces the observations equally well as the model displayed in Fig.~\ref{fig:L1527_HCO+model}. We can break this degeneracy using the \htcop observations. As shown in Fig.~\ref{fig:L1527_H13CO+model}, the \htcop emission at velocities $\geq |1.9|$ km s$^{-1}$ can be reproduced by a model with a constant \htcop abundance of $3\times10^{-12}$ in both disk and envelope. For an elemental $^{12}$C/$^{13}$C ratio of 68 \citep{Milam2005}, an \htcop abundance of $3\times10^{-12}$ suggests an \hcop abundance of $2\times10^{-10}$. Together these modeling results thus suggest that the \hcop abundance is lower in the disk than in the envelope, with abundances of $2\times10^{-10}$ in the outer disk ($>$ 60 au), $2\times10^{-11}$ at 40--60 au, and $\leq 2\times10^{-11}$ at radii $<$ 40 au. 

\begin{figure*}
\centering
\subfloat{\includegraphics[trim={0.2cm 17.2cm 7.7cm 0cm},clip]{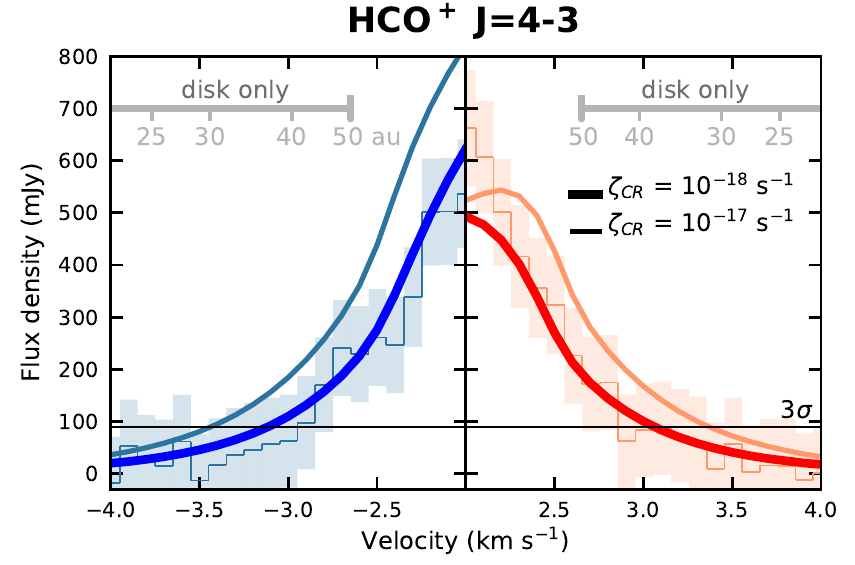}}
\subfloat{\includegraphics[trim={0.2cm 17.2cm 7.7cm 0cm},clip]{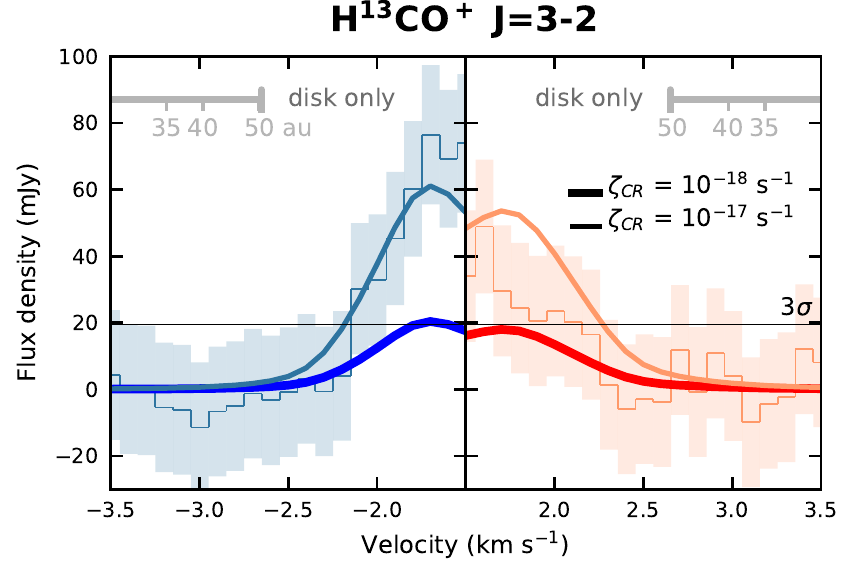}}
\caption{Spectra of the \hcop (left panel) and \htcop (right panel) emission extracted in a 0$\farcs$75 aperture centered on the blueshifted emission peak (left sides of each panel) and on the redshifted emission peak (right sides of each panel). The observations are displayed in discrete velocity bins of 0.1 \kms with the shaded area depicting the 3$\sigma$ uncertainty and a 10\% flux calibration uncertainty. The smooth lines are for models with a CO abundance of 10$^{-4}$ and a H$_2$O abundance of 10$^{-6}$, but with varying cosmic ray ionization rates of 10$^{-18}$ s$^{-1}$ (thick line; referred to as the fidicial model) and 10$^{-17}$ s$^{-1}$ (thin line; canonical cosmic ray ionization rate). The horizontal black line marks the 3$\sigma$ level. The velocity range containing only emission from the disk is marked by grey bars in the top of the panels, and the maximum radius probed at certain velocities is indicated. The displayed velocity range is different for both molecules.}
\label{fig:Spectra_fiducial-CR}
\end{figure*}

\citet{Jorgensen2004a} derived an \htcop abundance of $8.5\times10^{-12}$ for the envelope around L1527 from multiple single-dish observations. This is within a factor of three of our derived abundance of $3\times10^{-12}$, and consistent with our result that the abundance increases at larger radii. Our derived \hcop abundances on disk scales are consistent with the modeling results from \citet{Leemker2021} for protoplanetary disks around Herbig stars, which show a similar \hcop abundance gradient in the outer disk (20--100 au) and a stronger decrease across the snowline (4.5 au) with the abundance dropping below $10^{-12}$. The current observations are not sensitive enough to constrain such low abundances. However, since the \hcop abundance also depends on density and ionization, chemical modeling using a physical model specific for L1527 is required to link the abundance structure to the snowline.

\begin{figure*}
\centering
\includegraphics[width=\textwidth,trim={.5cm 9.6cm .3cm 2.3cm},clip]{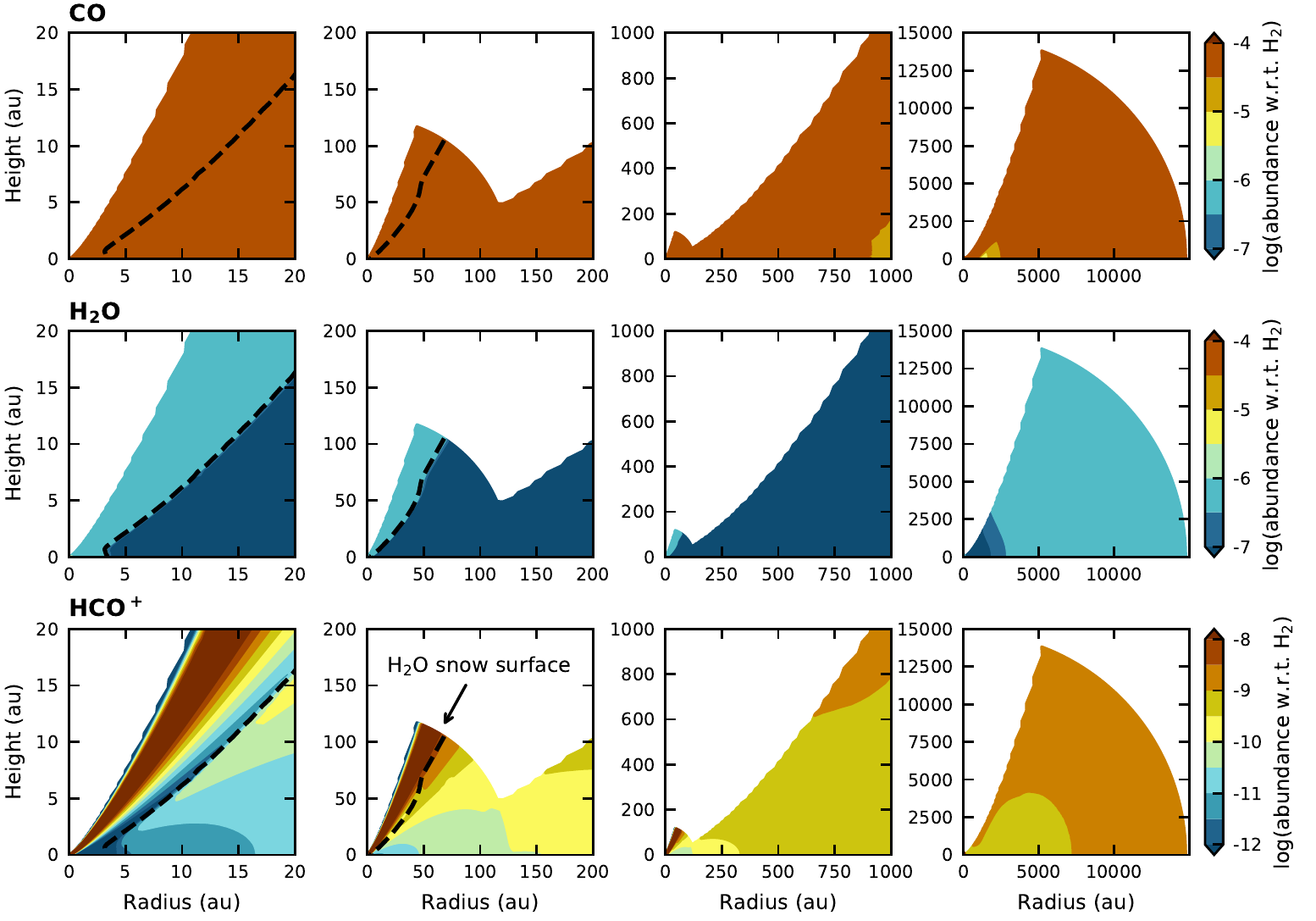}
\caption{Abundance structure for CO (top panels), H$_2$O (middle panels) and \hcop (bottom panels) predicted by the fiducial chemical model with initial CO and H$_2$O abundances of 10$^{-4}$ and 10$^{-6}$, respectively, and a cosmic ray ionization rate of $10^{-18}$ s$^{-1}$ for the physical structure of L1527. From left to right, panels display larger spatial scales. The disk outer radius is 125 au. The dashed line in the two left most columns marks the H$_2$O snow surface and the midplane snowline at 3.4 au.}
\label{fig:Chemistry_fiducial}
\end{figure*}


\subsection{Chemical modeling} \label{sec:ChemicalModel}

The chemical network presented by \citet{Leemker2021} was developed to study the relationship between the water snowline and \hcop (and \htcop) emission in Herbig disks. It contains the main formation and destruction reactions for \hcop as well as the freeze out and thermal desorption of water, as illustrated in Fig.~\ref{fig:ChemicalNetwork}. Reaction rate constants are taken from the UMIST database \citep{McElroy2013}, and are listed in Table 2 of \citet{Leemker2021}. For water, a binding energy of 5775 K is used, which corresponds to an amorphous water ice \citep{Fraser2001}. 

Freeze out of CO, the parent gas-phase molecule of \hcop, was not included in the study by \citet{Leemker2021} as this only occurred in the low-density outer region of the Herbig disks. Although there is no sign of CO freeze out in the disk around L1527 \citep{vantHoff2018b,vantHoff2020}, we have added the freeze out and thermal desorption of CO to the chemical network for completeness and to display the \hcop abundance structure in the envelope. The extact temperature at which CO desorbs depends on the composition of the ice \citep[e.g.,][]{Collings2003}, with pure CO ice desorbing at lower temperatures (855 K; \citealt{Bisschop2006}) than CO ice on top of water ice (1150 K; \citealt{Garrod2006}). The resulting desorption temperatures differ by $\sim$6 K; for example, 18 K versus 24 K for a density of $10^{7}$ cm$^{-3}$. In either case the CO snowline is located outside the L1527 disk, at $\sim$500 au or $\sim$200 au, respectively, and we adopt the binding energy of 855 K for a pure CO ice \citep{Bisschop2006}. Including or excluding freeze out of CO does not influence the \hcop emission in the disk and inner envelope velocity channels that we are interested in here. 

The freeze out rates depend on the available surface area of the dust grains. Following \citet{Leemker2021}, we assume a typical grain number density of $10^{12} \times n$(H$_2$) and a grain size of 0.1 $\mu$m. Even if a fraction of the grains have grown to larger sizes, the smallest grains will dominate the surface area and \citet{vantHoff2017} showed that adopting a more detailed description for the available surface area did not significantly affect the predicted N$_2$H$^+$ abundance for the protoplanetary disk around TW Hya. For the radiative transfer, we use the dust opacities from \citet{Tobin2013} as described at the beginning of Sect.~\ref{sec:Modeling}. 

Initially, all abundances are set to zero, except for H$_2$, gas-phase CO and gas-phase H$_2$O, and we run the chemistry for $10^5$ year. Running the chemistry for $10^6$ year, as typically done for protoplanetary disk studies, does not affect the \hcop abundance structure in the disk and inner envelope (radii $\lesssim$ 1000 au; see Appendix \ref{ap:Chemistry}). The main free parameters in the model are the initial CO and H$_2$O abundance and the cosmic ray ionization rate, which initiates the ion-neutral chemistry by ionizing H$_2$. The model does not include isotope-specific reactions and we adopt a $^{12}$C/$^{13}$C ratio of 68 \citep{Milam2005} to generate \htcop image cubes. 

\citet{vantHoff2018b} did not find evidence for a CO abundance much lower than the canonical value of $10^{-4}$ in the L1527 disk and \citet{Harsono2020} derived an upper limit for the H$_2$O abundance of $10^{-6}$. A model with these initial abudances reproduces the \hcop and \htcop observations equally well as the parametrized model described in Sect.~\ref{sec:ParametrizedModel}. For \hcop a cosmic ray ionization rate of $10^{-18}$ s$^{-1}$ (about one order of magnitude below the canonical value) needs to be adopted (see Fig.~\ref{fig:Spectra_fiducial-CR}). The asymmetry between blueshifted and redshifted emission is due to the kinematics of the disk and envelope, with the envelope in front of the disk for redshifted emission and the envelope behind the disk for blueshifted emission (see Appendix~\ref{ap:Chemistry}). Since the \hcop observations are more sensitive to the disk than the \htcop observations, as discussed in Sections \ref{sec:Observations} and \ref{sec:ParametrizedModel}, we will focus first on the model that reproduces the \hcop emission and we will discuss the ionization rate in more detail in Sect.~\ref{sec:CRrate}.

The abundance structure reproducing the \hcop observations (our fiducial model with X(CO) = 10$^{-4}$, X(H$_2$O) = 10$^{-6}$ and $\zeta_{\rm{CR}} = 10^{-18}$ s$^{-1}$) is presented in Fig.~\ref{fig:Chemistry_fiducial}. For the adopted temperature and density structure, the water snowline is located at 3.4 au, which corresponds to a temperature of 140 K. The snow surface is located high up in the disk surface layers, making most of the disk and the inner envelope devoid of gas-phase water. Water is present in the gas phase at radii $\gtrsim$ 3000 au because the density in the outer envelope becomes too low for water to freeze out in the timescale of the model. A similar water abundance profile was found by \citet{Schmalzl2014}, who used a simplified chemical network that was benchmarked against three full chemical networks to model \textit{Herschel} observations of water in protostellar envelopes. 

Overall, the \hcop abundance gradually decreases with increasing density and steeply decreases across the water snow surface. At the high midplane densities, the \hcop abundance remains low directly outside the water snowline as shown in earlier work \citep{vantHoff2018a,Hsieh2019}. In the midplane, the \hcop abundance drops from $10^{-11}$ at 16 au to $3\times10^{-12}$ at 5 au and then steeply drops to abundances $< 10^{-12}$ inside the snowline. These abundances are all at least an order of magnitude lower than the upper limit derived for the high velocity channels probing radii $\leq 40$ au using the parametrized model in Sect.~\ref{sec:ParametrizedModel}. The sensitivity of the observations is thus not high enough to probe the \hcop abundance drop across the snowline and the absence of emission in the highest velocity channels cannot be linked to the snowline. The high \hcop abundance in the uppermost surface layers of the disk is likely because CO photodissociation is not included in the model. In this region, the rate for the reaction between \hcop and H$_2$O, $R_\mathrm{destruction}$,
\begin{equation}
R_{\mathrm{destruction}} \propto n(\mathrm{HCO}^+) n(\mathrm{H}_2\mathrm{O}), 
\end{equation}
is low, because the low density results in a low H$_2$O number density, $n$(H$_2$O). As discussed by \citet{Leemker2021}, electron recombination becomes the dominant destruction mechanism of \hcop in this region. At the same time, the \hcop formation rate, $R_\mathrm{formation}$,
\begin{equation}
R_{\mathrm{formation}} \propto n(\mathrm{CO}) n(\mathrm{H}_3^+), 
\end{equation}
remains high as the H$_3^+$ number density is set by the cosmic ray ionization rate and is therefore independent of density. Including CO photodissociation would remove the parent molecule CO and hence prevent \hcop formation, but knowledge of the UV field is required for a proper implementation. However, this low-density layer does not significantly contribute to the total \hcop emission. Manually removing this layer before radiative transfer results in flux differences less than 0.2\% and identical spectra as to those displayed in Fig.~\ref{fig:Spectra_fiducial-CR}. 

The \hcop abundance is barely influenced by CO freeze out, because this occurs only in a small region of the envelope. In the disk and inner envelope ($\lesssim$900 au), the temperature is too high for CO to freeze out, while at radii $\gtrsim$2500 au the density becomes too low for CO to freeze out within $10^5$ yr, resulting in CO being present in the gas phase at the initial abundance throughout the majority of the system. A similar abundance profile was derived by \citet{Jorgensen2005} for a sample of 16 sources, and $^{13}$CO has been observed out to radii of $\sim$10,000 au toward L1527 \citep{Zhou1996}. Running the chemistry for $10^6$ yr results in a midplane region with a decreased \hcop abundance centered around 2500 au due to higher levels of CO freeze out. Simultaneously, the higher levels of H$_2$O freeze out increase the \hcop abundance, resulting overall in only a small region with a lower \hcop abundance at radii $\gtrsim$1000 au after $10^6$ year. The \hcop abundance at smaller radii is unaffected (see Fig.~\ref{fig:Chemistry_1e6yr}).

\begin{figure*}
\centering
\subfloat{\includegraphics[trim={0.2cm 17.2cm 7.7cm 0cm},clip]{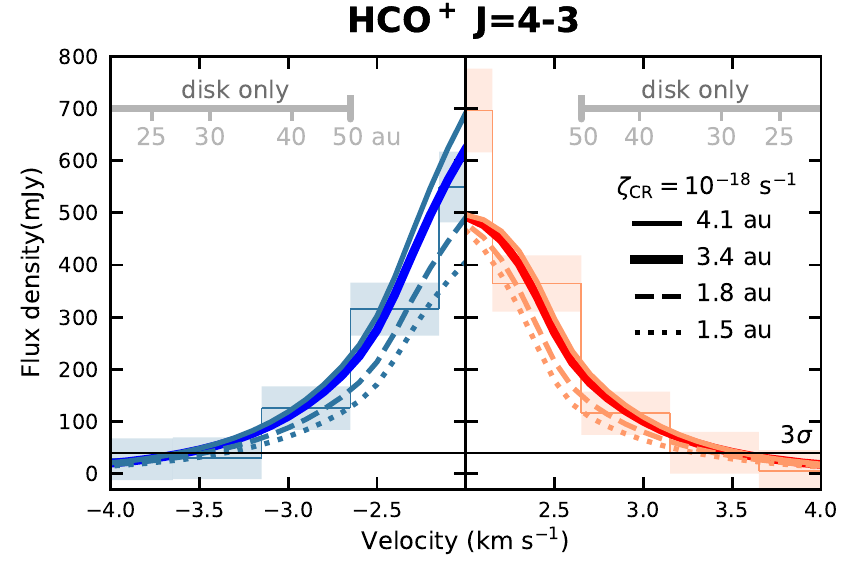}}
\subfloat{\includegraphics[trim={0.2cm 17.2cm 7.7cm 0cm},clip]{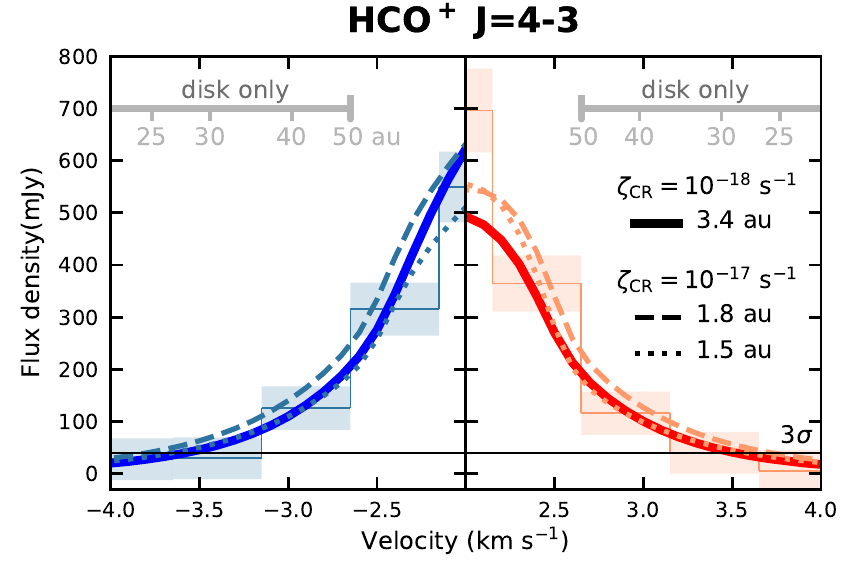}}
\caption{Spectra of the \hcop emission extracted in a 0$\farcs$75 aperture centered on the blueshifted emission peak (left side of each panel) and on the redshifted emission peak (right side of each panel). The observations are binned to 0.5 km s$^{-1}$ and the shaded area depicts the 3$\sigma$ uncertainty and a 10\% flux calibration uncertainty. The thick dark colored lines are for the fiducial model with a snowline at 3.4 au. The other lines represent models where the disk temperature has been multiplied by a constant factor to obtain snowline locations of 1.5 au (dotted line), 1.8 au (dashed line) and 4.1 au (thin solid line). The cosmic ray ionization rate is $10^{-18}$ s$^{-1}$ for the models in the left panel and $10^{-17}$ s$^{-1}$ in for the models in the right panel, except for the fiducial model (thick line) which is the same as in the left panel. The horizontal black line marks the 3$\sigma$ level. The velocity range containing only emission from the disk is marked by grey bars in the top of the panels, and the maximum radius probed at certain velocities is indicated. }
\label{fig:Spectra_Tdisk}
\end{figure*}

The \hcop abundance structure in the disk as presented in Fig.~\ref{fig:Chemistry_fiducial} is very robust with respect to changes in the initial CO and H$_2$O abundance.  \citet{Aso2017} suggested a CO abundance of $2\times10^{-5}$ based on ALMA observations of C$^{18}$O, but lowering the CO abundance two orders of magnitude does not affect the \hcop abundance in the disk significantly (see Fig.~\ref{fig:Chemistry_lowCO}). The only changes are a lower \hcop abundance in the upper most surface layers of the disk (above the snow surface) and a lower abundance at radii $\gtrsim$900 au. A canonical H$_2$O abundance of $10^{-4}$ increases the vertical height of the \hcop depleted layer above the snow surface and strongly decreases the \hcop abundance at radii $\gtrsim$1500 au (see Fig.~\ref{fig:Chemistry_highH2O}). Lowering the H$_2$O abundance further to $10^{-7}$ slightly increases the \hcop abundance above the snow surface. The \hcop abundance thus remains unaltered throughout the majority of the disk for CO abundances between $10^{-4}$ and $10^{-6}$ and H$_2$O abundances between $10^{-4}$ and $10^{-7}$. 

The only parameter that has a strong impact on the \hcop abundance in the disk is the cosmic ray ionization rate. As described in \citet{vantHoff2018a}, the \hcop abundance is proportional to the square root of the cosmic ray ionization rate. Hence, a canonical value of $10^{-17}$ s$^{-1}$ results in a \hcop abundance higher by a factor $\sim$3 compared to a rate of $10^{-18}$ s$^{-1}$ (Fig.~\ref{fig:Chemistry_CR} versus Fig.~\ref{fig:Chemistry_fiducial}), and a too high \hcop flux compared to what is observed (see Fig.~\ref{fig:Spectra_fiducial-CR}). The predicted \hcop flux for a CR ionization rate of $10^{-17}$ s$^{-1}$ is less than a factor 3 higher than the flux predicted for a rate of $10^{-18}$ s$^{-1}$, signaling that the emission becomes optically thick.


\section{Discussion}


\subsection{The water snowline location in L1527}\label{sec:SnowlineLocation}

For the temperature and density structure derived by \citet{Tobin2013}, the water snowline is predicted to be at 3.4 au in the disk around L1527 and the corresponding \hcop abundance structure from a small chemical network calculation can reproduce the observed \hcop emission. Although the current observations are not sensitive to the expected \hcop abundance changes across the snowline, the chemical model shows a decrease in the \hcop abundance over a much larger radial range above the snow surface. This suggests that the snowline location may be constrained indirectly from \hcop emission based on the global temperature structure. To investigate the stringency of the current observations, we run a set of models with different snowline locations separated by $\sim$0.2--0.5 au generated by multiplying the fiducial temperature structure in the disk by a constant factor. To obtain the maximum sensitivity, we bin the observations to 0.5 km s$^{-1}$. 

Figure~\ref{fig:Spectra_Tdisk} displays the \hcop spectra for models with a snowline at 1.5, 1.8, 3.4 (fiducial) and 4.1 au. The differences between these models are too small to be distinguished at disk-only velocities, but the current sensitivity is high enough to see a significant difference at a $\pm$2.4 \kms velocity offset. While a snowline at 1.8 au produces \hcop emission at about the 3$\sigma$ uncertainty level (including a 10\% flux calibration uncertainty), a snowline at 1.5 au clearly underproduces the emission. The effect of a warmer disk is not very pronounced at redshifted velocities, but a snowline at 4.1 au slightly overproduces the blueshifted emission. Lower \hcop emission as a result of a colder disk can partially be compensated by a higher cosmic ray ionization rate. For $\zeta_{\rm{CR}} = 10^{-17}$ s$^{-1}$, a snowline at 1.5 au can reproduce the observed \hcop emission, while a snowline at 1.8 au results in a too high flux. However, \citet{vantHoff2018b} was not able to reproduce the $^{13}$CO and C$^{18}$O $J=2-1$ emission with the temperature structure corresponding to a 1.5 au snowline (their Intermediate Model), instead requiring a warmer disk. Changing the temperature also in the envelope has only a small effect on the emission at $\pm$2.4 \kms (Fig.~\ref{fig:Spectra_Tdisk-full}). Taken together, for the here adopted physical structure, these results thus suggest a snowline radius of 1.8-4.1 au if $\zeta_{\rm{CR}} = 10^{-18}$ s$^{-1}$ and between 1.5 and 1.8 au for $\zeta_{\rm{CR}} = 10^{-17}$ s$^{-1}$. 

No other molecular line observations are currently available to locate the water snowline in L1527: H$_2^{18}$O emission has not been detected \citep{Harsono2020}, and while weak methanol emission has been observed, the sensitivity and resolution of that data were insufficient to determine its spatial origin \citep{Sakai2014b}. \citet{Aso2017} presented a warmer model based on fitting of sub-millimeter continuum visibilities than the model derived by \citet{Tobin2013} and used in this work, with the snowline at $\sim$8 au. This is at least twice as far out as predicted here based on the \hcop models, but the temperature profile is kept fixed in the fitting procedure of \citet{Aso2017}. \citet{vantHoff2018b} inferred a temperature profile based on optically thick $^{13}$CO and C$^{18}$O emission, which has a temperature of $\sim$35 K at 40 au. The temperature in the inner $\sim$20 au depends strongly on the chosen power-law coefficient and does not provide a strong constraint on the snowline location. 

\begin{figure}
\centering
\includegraphics[trim={0.2cm 16.5cm 7.7cm 0.2cm},clip]{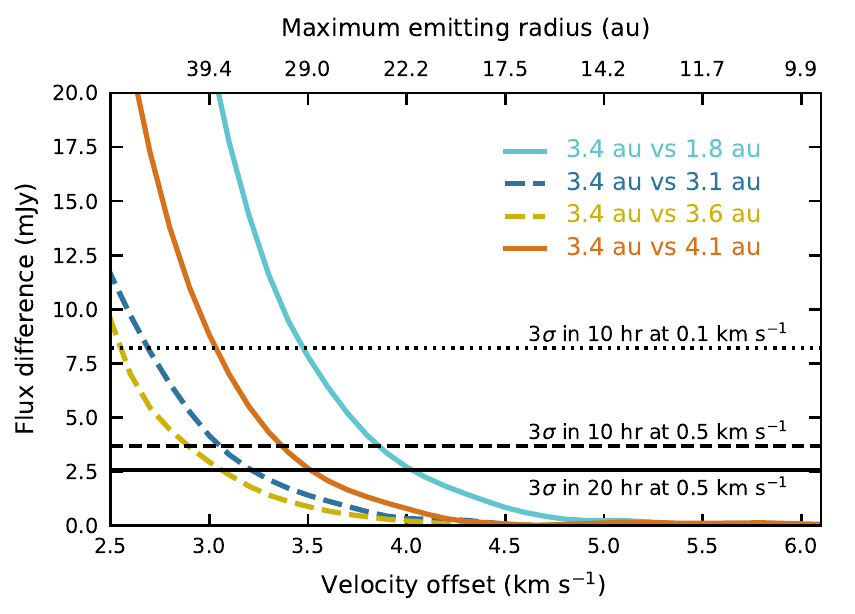}
\caption{Difference in \hcop $J=4-3$ flux between models with different snowline locations compared to the fiducial model with the snowline at 3.4 au at velocity offsets that trace only the disk. The top horizontal axis shows the maximum radius probed at a certain velocity for an edge-on Keplerian disk around a 0.4 $M_\odot$ star as used for L1527. The dotted and dashed lines mark the 3$\sigma$ limit that can be reached with ALMA in 10 hours on source at a spectral resolution of 0.1 and 0.5 km s$^{-1}$, respectively. The solid line is the 3$\sigma$ limit at 0.5 km s$^{-1}$ for 20 hours on source.  The 3$\sigma$ sensitivity of the current \hcop observations is 40 mJy at 0.5 km s$^{-1}$.}
\label{fig:RequiredSensitivity}
\end{figure}

Providing stronger constraints on the snowline location using \hcop emission will require significantly deeper observations as illustrated in Fig.~\ref{fig:RequiredSensitivity}. As higher velocities trace emission originating out to smaller radii, the total flux decreases at higher velocities, and hence higher sensitivity is required to distinguish two models. Ideally, one would want to compare the flux in a certain velocity channel with the flux of models with the snowline inside and outside the maximum radius probed by that velocity. However, it is immediately clear from Fig.~\ref{fig:RequiredSensitivity} that this is not possible for L1527 with current facilities as flux differences at $\gtrsim |4.0|$ km s$^{-1}$ (corresponding to a $\sim$20 au radius) become too small to be observed in 20 hours on source with ALMA. 

Nonetheless, deeper integrations will allow for better constraints in two ways. First, a higher sensitivity will allow different models to be compared over more velocity channels, and in particular, in channels that only trace disk emission. This will remove any influence from the envelope. For example, 10 hours on source with ALMA will result in 5--10 0.1 km s$^{-1}$ disk-only channels (or 2--3 channels at 0.5 km s$^{-1}$) where models with snowlines at 1.8, 3.4 and 4.1 au can be distinguished, as compared to currently one 0.5 km s$^{-1}$ channel containing emission from both disk and envelope. Second, a higher sensitivity will allow to distinguish between models with smaller differences in snowline radius. With 10 hours on source, the snowline can be constrained within a few tenths of an au, although there will be a degeneracy with the cosmic ray ionization rate. 

Higher \textit{J} transitions will generally trace warmer and denser material, but in turn, higher dust opacity at higher frequencies may prevent to observe these transitions from the inner disk. For the here adopted dust opacities, the continuum $\tau=1$ surface shift outward by only $\sim$1 au for the $J=8-7$ transition (713.342090 GHz) compared to the $J=4-3$ transition (356.734288 GHz) (Fig.~\ref{fig:VelocityField}), suggesting that the dust opacity is not a strong limiting factor in choosing an \hcop transition. For proper treatment of higher $J$ transitions, UV heating of the gas has to be taken into account and this emission may arise from a thin layer in which the gas and dust temperatures are decoupled \citep[e.g.,][]{Visser2012}. Such a modeling approach was adopted by \citet{Leemker2021}, while we here adapt the dust temperature structure for L1527 and assume the gas and dust temperature are equal, which is appropriate for the $J=4-3$ transition which emits from regions where the temperatures are coupled \citep[e.g.,][]{Mathews2013}. 

That being said, in our model the integrated flux at high velocities ($\gtrsim |3|$ \kms) increases up to a factor $\sim$2 or $\sim$3 for the $J=5-4$ (445.902996 GHz) and $J=7-6$ (624.208673 GHz) transitions, respectively, compared to the $J=4-3$ flux in the fiducial model (not shown). The flux of the $J=8-7$ transition is similar to the $J=7-6$ flux at velocities originating solely in the disk. For the $J=7-6$ transition, the curves in Fig.~\ref{fig:RequiredSensitivity} shift to the right by $\sim$1 \kms, suggesting that it is easier to distinguish between different snowline locations. However, the decrease in atmospheric transmission results in significantly lower sensitivities that make high sensitivity observations very expensive: in 20 hours on source at 0.5 \kms resolution, an rms of 224 mJy, 25 mJy and 20 mJy is reached for $J=5-4$, $J=7-6$, and $J=8-7$, respectively. These sensitivities would just be enough to distinguish between a snowline at 1.8, 3.4 or 4.1 au at velocity offsets $<$ 3 \kms with the $J=7-6$ and $J=8-7$ transitions. As even for the $J=1-0$ (89.188523 GHz) transition the snowline coincides with the dust $\tau=1$ surface, the \hcop $J=4-3$ transition is the best suited to constrain the snowline location. 

With such long integration times required to derive stronger constraints on the snowline location from \hcop emission, it is worth investigating whether the snowline can be located directly with water observations. As shown in Fig.~\ref{fig:VelocityField}, the snowline is expected to be hidden behind optically thick dust for frequencies above $\sim$90 GHz, so a direct detection of the snowline would not be possible for L1527. However, locating the snowline directly with water observations may still turn out to be a viable route for sources that have the water snowline extend beyond the radius where the dust becomes optically thick.

\subsection{Cosmic ray ionization rate}\label{sec:CRrate}

Another important result concerns the cosmic ray ionization rate in the disk of L1527. The cosmic ray ionization rate is crucial for both the physical and chemical evolution of the disk. From a physical perspective, ionization plays an important role in angular momentum transport through the magneto-rotational instability (MRI; e.g., \citealt{Balbus1991}). From a chemical point of view are cosmic rays the driver of chemistry in the disk midplane, where other ionizing agents cannot penetrate \citep[e.g.,][]{Eistrup2016,Eistrup2018}. For the here adopted physical structure of L1527 that is able to reproduce multi-wavelength continuum emission \citep{Tobin2013} as well as CO isotopologue emission \citep{vantHoff2018b}, a canonical CR ionization rate of $10^{-17}$ s$^{-1}$ overproduces the \hcop emission which originates predominantly from radii $\gtrsim$40 au (Fig.~\ref{fig:Spectra_fiducial-CR}). The \htcop emission, which is originating predominantly from the inner envelope, does require a CR ionization rate of $10^{-17}$ s$^{-1}$ (Fig.~\ref{fig:Spectra_fiducial-CR}). 

In order to reproduce the \hcop observations with a CR ionization rate of $10^{-17}$ s$^{-1}$ the disk temperature needs to be lowered such that the snowline is located inside of 1.8 au (instead of 3.4 au; Fig.~\ref{fig:Spectra_Tdisk}). A temperature structure obtained by multiplying our fiducial temperature with a constant factor of 0.6, resulting in a snowline at 1.5 au, is too cold to explain the $^{13}$CO and C$^{18}$O $J=2-1$ emission \citep{vantHoff2018b}, but the CO isotopologue observations are not sensitive enough to confidently say whether the temperature structure associated with a 1.8 au snowline is too cold as well. We have assumed that the temperature changes globally by a constant factor, and this analysis cannot rule out that a model with a slightly flatter temperature profile (i.e., colder in the inner few au) would be able to explain the molecular line observations with a canonical CR ionization rate. Higher sensitivity observations of CO isotopologues or other temperature tracers are required to better constrain the detailed temperature structure. 

A snowline location different from 3.4 au could be obtained by, for example, a different luminosity or different disk mass. Measurements of the bolometric luminosity based on the spectral energy distribution (SED) range between 1.6 and 2.0 $L_\odot$ \citep{Tobin2008,Green2013,Karska2018}. This is likely an underestimation of the true luminosity as edge-on sources embedded in an envelope can have internal luminosities up to $\sim$2 times higher than the bolometric luminosity \citep{Whitney2003}. For a 1 $L_\odot$ protostar, \citet{Tobin2008} require an accretion luminosity of 1.6 $L_\odot$ to fit the SED, resulting in a true bolometric luminosity of $\sim$2.6 $L_\odot$. The model used here has a total luminosity of 2.75 $L_\odot$ \citep{Tobin2013}, and assuming that the snowline radius scales as the square root of the luminosity, a luminosity of 0.8 $L_\odot$ would be required for a snowline radius of 1.8 au. This is a factor two smaller than derived from the SED. A lower disk mass could also shift the snowline inward, but for an accretion rate of $3\times10^{-7} M_\odot$ yr$^{-1}$ (corresponding to an accretion luminosity of 1.6 $L_\odot$), models by \citet{Harsono2015} show less than 1 au difference between disk masses of 0.05 and 0.1 $M_\odot$. The disk mass of the model used here is 0.0075 $M_\odot$. Modeling of high resolution ALMA data, for example, from the FAUST or eDisk large programs may provide additional constraints on the disk structure.  

Chemically, a canonical CR ionization rate may be reconciled with the observations if there is a higher destruction rate of \hcop. In our model, \hcop can be destroyed by H$_2$O and electrons, where the electrons are provided by ionization of H$_2$ by cosmic rays. Since grains are likely negatively charged \citep{Umebayashi1980}, ions may also recombine on dust grains. The recombination rate for this process, $R_{\mathrm{grain}}$, is given by 
\begin{equation}
R_{\mathrm{grain}} = a_gn_\mathrm{H}n(\mathrm{HCO}^+), 
\end{equation}
where $a_g$ is the recombination rate coefficient ($a_g \approx 10^{-17}$ cm$^{3}$ s$^{-1}$; \citealt{Umebayashi1980}), and $n_\mathrm{H}$ and $n(\mathrm{HCO}^+)$ are the hydrogen and \hcop number density. The recombination rate in the gas phase, $R_{\mathrm{gas}}$, is given by 
\begin{equation}
R_{\mathrm{gas}} = kn_\mathrm{e}n(\mathrm{HCO}^+),
\end{equation}
where $n_{\mathrm{e}}$ is the eletron density. The reaction rate coefficient, $k$, has a temperature dependence, and is $4-8 \times 10^{-7}$ cm$^{3}$ s$^{-1}$ for temperatures between 150 and 50 K (UMIST database; \citealt{McElroy2013}). This means that the grain recombination rate becomes larger than the gas-phase recombination rate for electron abundances, $n_\mathrm{e}/n_\mathrm{H}$, $\gtrsim$10$^{-11}$. Since the electron abundance is approximately equal to the \hcop abundance, this condition is only met in the disk midplane inside $\sim$16 au for the fiducial model (Fig.~\ref{fig:Chemistry_fiducial}) and only inside $\sim$5 au for the model with a CR ionization rate of $10^{-17}$ s$^{-1}$ (Fig.~\ref{fig:Chemistry_CR}). Destruction of \hcop via electron recombination on grains is thus unlikely to effect the predicted \hcop abundance.  

While we cannot fully rule out a canonical CR ionization rate, the different ionization rates derived from \hcop and \htcop are not necessarily in conflict with each other. The lower $J$ transition and lower velocities probed with \htcop make the \htcop observations more sensitive to the inner envelope than the \hcop observations. This would then suggest that the CR ionization rate is lower in the disk compared to the envelope, which could simply be the result of stronger attenuation of external cosmic rays due to the higher density of the disk. The cosmic ray ionization rate is expected to decrease exponentially with an attenuation column of 96 g cm$^{-2}$ \citep{Umebayashi1981,Umebayashi2009} or even higher \citep{Padovani2018}. However, a column larger than 96 g cm$^{-2}$ is only reached in the inner 0.5 au in our L1527 model. Another explanation for a low cosmic ray ionization rate in the disk may be the exclusion of cosmic rays by stellar winds and/or magnetic fields as proposed by \citet{Cleeves2015} for the protoplanetary disk around TW Hya. The same mechanism could explain the gradient in cosmic ray ionization rate derived for the IM Lup disk, where the steep increase in CR ionization rate in the outer disk may indicate the boundary of the ``T Tauriosphere'', that is, a stellar-wind-induced boundary analogous to the Sun's heliosphere \citep{Seifert2021}.

While models show that cosmic rays can be produced by jet shocks and by accretion shocks at protostellar surfaces \citep{Padovani2015,Padovani2016}, the transport of cosmic rays in protostellar disks is very complicated (as shown for external CRs by \citealt{Padovani2018}). Models by \citet{Gaches2018} for the simpler case of protostellar cores show five orders of magnitude difference in CR ionization rate between the two limiting cases of transport of internally created CRs through the core. 

In our chemical network, all ionization is provided by cosmic rays, but UV and X-ray ionization can play a role as well, in particular in higher layers in the disk with X-rays penetrating deeper than UV radiation (see e.g., \citealt{Cleeves2014,Notsu2021,Seifert2021}). However, it is not clear at what point during stellar evolution the dynamo turns on and X-rays are emitted, and no X-ray emission has been detected toward L1527 with \textit{Chandra} \citep{Gudel2007}. Since the observations constrain the \hcop abundance, a significant contribution of UV and/or X-rays to the \hcop chemistry would mean that the cosmic ray ionization rate is even lower than $10^{-18}$ s$^{-1}$.

Other observational constraints for the CR ionization rate in the L1527 disk do not currently exist. \citet{Favre2017} used \textit{Herschel} observations of the ratio between \hcop $J=6-5$ and N$_2$H$^+$ $J=6-5$ to constrain the ionization rate in Class 0 protostars, but the upper limit resulting from the non-detection of N$_2$H$^+$ toward L1527 only constrains the CR ionization rate to be smaller than $10^{-14}$ s$^{-1}$. The signal-to-noise ratio of the $J=6-5$ observations is too low to detect emission at velocity offsets $\gtrsim$2 \kms, so they do not help in constraining the \hcop distribution or disk temperature structure.   

If confirmed, a low CR ionization rate in a young disk may have profound consequences as high ionization levels are crucial for disk evolution. For example, for angular momentum transport through MRI, the gas needs to be coupled to the magnetic field \citep{Gammie1996}, and hence insufficient ionization may suppress MRI and create a low-turbulence ``dead zone'', favorable for planetesimal formation \citep[e.g.,][]{Gressel2012}. From a chemical perspective, the up to two orders of magnitude CO depletion observed in protoplanetary disks can only be reproduced by models with CR ionization rates on the order of $10^{-17}$ s$^{-1}$ \citep{Bosman2018,Schwarz2019}. Currently the CO snowline is located outside the L1527 disk \citep{vantHoff2018b}, but unless the CR ionization rate would increase at later stages, a CR ionization rate of $10^{-18}$ s$^{-1}$ would suggest that no chemical processing of CO will occur in the L1527 disk once the disk has cooled enough for the CO snowline to shift inward.


\section{Conclusions} \label{sec:Conclusions}

We have presented $\sim0\farcs5$ ($\sim$70 au) resolution ALMA observations of \hcop $J=4-3$ and \htcop $J=3-2$ toward the embedded disk L1527. In order to constrain, for the first time, the water snowline location in a young disk, we modeled the \hcop abundance and emission using a physical model specific for L1527 \citep{Tobin2013} and a small chemical network (based on \citealt{Leemker2021}). Our main results are summarized below. 

\begin{itemize}
\item The observed \hcop emission traces the disk down to a radius of $\sim$40 au. The emission can be reproduced with the L1527-specific physical structure that has the water snowline at 3.4 au, given that the cosmic ray ionization rate is lowered to $10^{-18}$ s$^{-1}$. 
\item Even though the observations are not sensitive to the expected \hcop abundance change across the midplane snowline, the change across the radial snow surface and the global temperature structure allow us to constrain the snowline location between 1.8 and 4.1 au by multiplying the fidicial temperature structure with a constant factor. The snowline can be inward of 1.8 au if the CR ionization rate is $10^{-17}$ s$^{-1}$, but a previous analysis showed that a temperature structure with the snowline at 1.5 au is too cold to reproduce the $^{13}$CO and C$^{18}$O observations.  
\item The \hcop abundance structure in the disk predicted by the small chemical network is very robust for the initial H$_2$O and CO abundance, and only significantly depends on the cosmic ray ionization rate. 
\item The observed \htcop emission extends out to lower velocity offsets than the \hcop emission, indicating that the emission predominantly originates in the inner envelope. For the adopted physical structure, a canonical CR ionization rate of $10^{-17}$ s$^{-1}$ is required to reproduce the \htcop emission. Together, the \hcop and \htcop results suggest that the CR ionization rate has a canonical value of $10^{-17}$ s$^{-1}$ in the inner envelope and may be attenuated to $\sim10^{-18}$ s$^{-1}$ in the disk. 
\end{itemize}

These results demonstrate the use of \hcop as a snowline tracer in embedded disks. However, as long integration times with ALMA are required to detect emission at high velocities to eliminate envelope contribution and to constrain the snowline to within 0.5 au, the direct detection of the snowline through observations of water isotopologues may still prove to be a viable strategy. Deep water observations of a range of different sources are required to constrain when water observations are viable and when we have to retort to indirect tracing with \hcop. Observations of water ice with the James Webb Space Telescope may provide constraints on the (vertical) snowline as well. In sources with a direct measurement of the snowline location, \hcop observations will allow to constrain the cosmic ray ionization rate.


\acknowledgments 

We would like to thank the referee, Ewine van Dishoeck, Martijn van Gelder, and Naomi Hirano for feedback on the manuscript. This paper makes use of the following ALMA data: ADS/JAO.ALMA\#2012.1.00193.S and ADS/JAO.ALMA\#2012.1.00346.S. ALMA is a partnership of ESO (representing its member states), NSF (USA) and NINS (Japan), together with NRC (Canada), MOST and ASIAA (Taiwan), and KASI (Republic of Korea), in cooperation with the Republic of Chile. The Joint ALMA Observatory is operated by ESO, AUI/NRAO and NAOJ. M.L.R.H acknowledges support from the Michigan Society of Fellows. M.L. acknowledges support from the Dutch Research Council (NWO) grant 618.000.001. J. J. T. acknowledges funding from NSF grant AST-1814762. The National Radio Astronomy Observatory is a facility of the National Science Foundation operated under cooperative agreement by Associated Universities, Inc. J.K.J. acknowledges support from the Independent Research Fund Denmark (grant No. 0135-00123B). E.A.B. acknowledges support from NSF AAG Grant \#1907653. Astrochemistry in Leiden is supported by the Netherlands Research School for Astronomy (NOVA).


\bibliography{References}{}
\bibliographystyle{aasjournal}




\restartappendixnumbering

\begin{appendix}


\section{Physical and kinematical structure of L1527}\label{ap:PhysicalStructure}

\begin{figure}
\centering
\subfloat{\includegraphics[trim={0cm 15.3cm 8cm 0cm},clip]{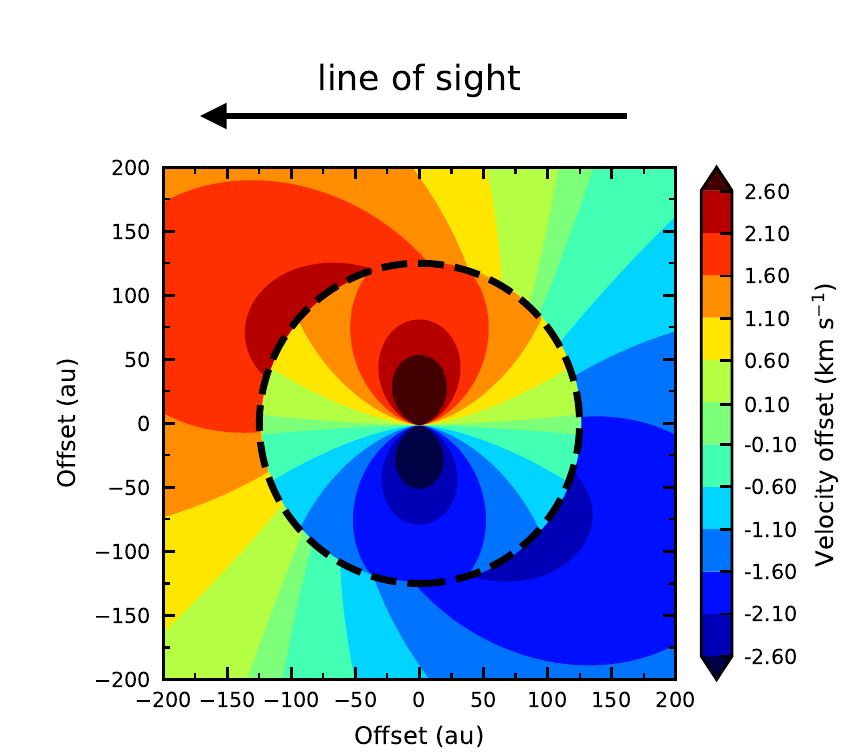}}
\subfloat{\includegraphics[trim={0cm 15.3cm 8cm 0cm},clip]{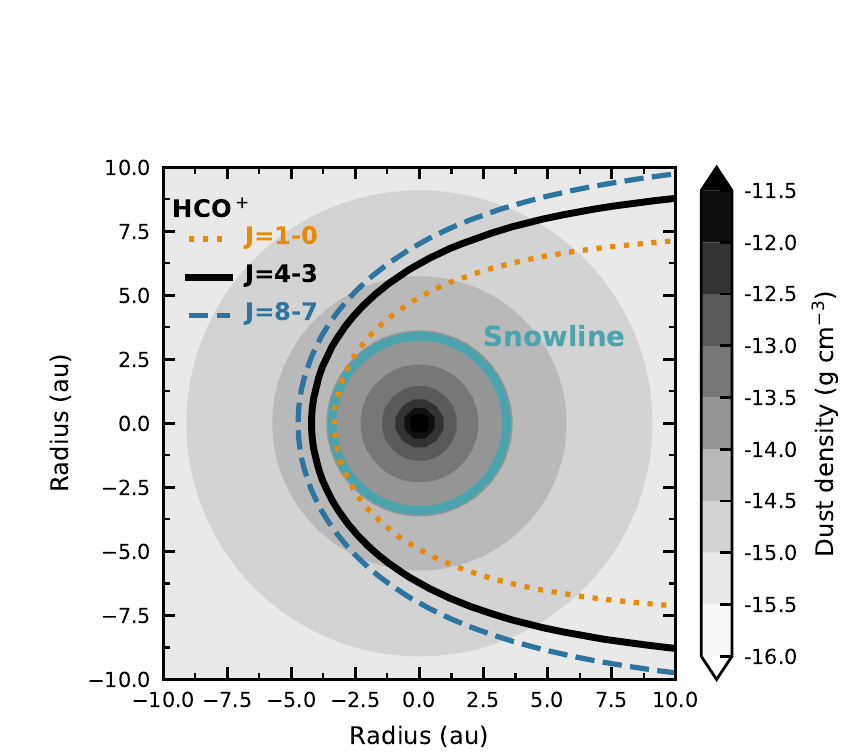}}
\caption{Left panel: Face-on view of the midplane in the disk and inner envelope of L1527, showing the velocity component along the line of sight (the observer is on the left as indicated by the black arrow above the panel). The material in the 125 au disk (marked by a dashed line) has a Keplerian velocity, while the envelope material has a rotating infalling velocity profile \citep{Ulrich1976,Cassen1981}. The adopted stellar mass is 0.4 $M_\odot$. Right panel: Dust density in the inner 10 au and the $\tau$ = 1 surfaces for the dust emission in our model at the wavelengths of the \hcop $J=1-0$ (89.188523 GHz; dotted orange line), $J=4-3$ (356.734288 GHz; solid black line) and $J=8-7$ (713.342090 GHz; dashed blue line) transitions. The snowline at 3.4 au is indicated by the blue circle.}
\label{fig:VelocityField}
\end{figure}

\begin{figure}
\centering
\includegraphics[trim={0.2cm 17.2cm 7.7cm 0cm},clip]{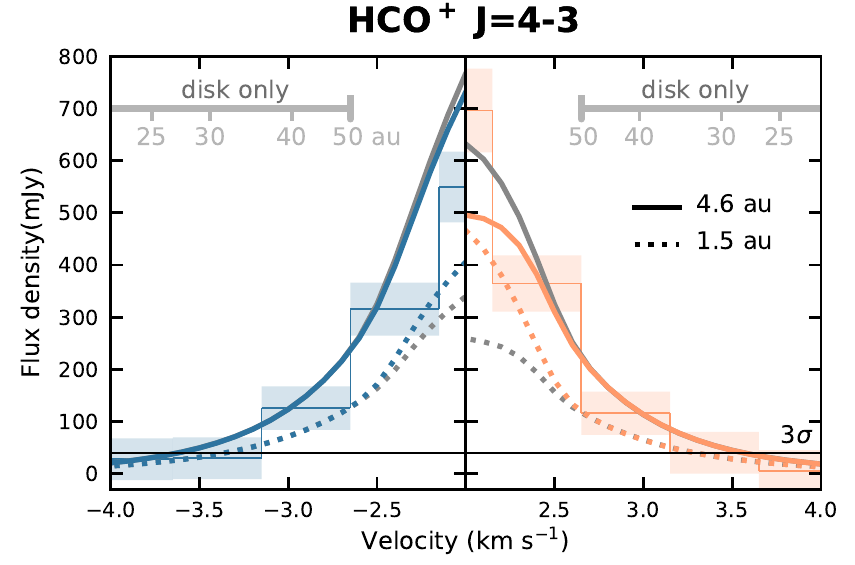}

\caption{Spectra of the \hcop emission extracted in a 0$\farcs$75 aperture centered on the blueshifted emission peak (left) and on the redshifted emission peak (right). The observations are binned to 0.5 km s$^{-1}$ and the shaded area depicts the 3$\sigma$ uncertainty and a 10\% flux calibration uncertainty. The solid lines are for models with a snowline at 4.6 au and the dotted lines are for models with a snowline at 1.5 au. For the colored lines, the temperature factor required for that particular snowline location has only been applied to the disk, while for the gray lines, this temperature factor has been applied to both disk and envelope. The difference between colored and gray lines is thus only the temperature in the envelope. The horizontal black line marks the 3$\sigma$ level. The velocity range containing only emission from the disk is marked by grey bars in the top of the panels, and the maximum radius probed at certain velocities is indicated. }
\label{fig:Spectra_Tdisk-full}
\end{figure}

Figure~\ref{fig:VelocityField} presents the velocity along the line of sight for the material in the disk and inner envelope midplane. The highest velocities are only reached in the inner disk, presenting an opportunity to study disk emission unaffected by the envelope. At intermediate velocities ($\Delta|v| \sim 1.6-2.6$ \kms), redshifted emission from the envelope originates in front of the disk, while at blueshifted velocities we see the disk in front of the envelope. This can result in asymmetric line profiles, as for example in Fig.~\ref{fig:Spectra_fiducial-CR}, especially when the emission becomes optically thick. For \hcop, we can illustrate this clearly by comparing models where the temperature modification as described in Sect.~\ref{sec:SnowlineLocation} is applied only to the disk or to both the disk and envelope. As can be seen in Fig.~\ref{fig:Spectra_Tdisk-full}, the blueshifted emission at velocities $\leq -2.0$ km s$^{-1}$ is hardly affected by a change in envelope temperature, signaling that most of the emission is originating in the disk. The redshifted emission is stronger affected and the redshifted emission at velocities $\leq 2.5$ km s$^{-1}$ thus has a strong envelope component. Another effect that comes in to play at lower redshifted velocities is absorption due to the foreground infalling envelope. This effect together with larger scale emission being resolved out may explain why the model better reproduces the blueshifted \htcop emission as the convolution of the synthetic image cube does not capture the effects of the interferometer. 

Figure~\ref{fig:VelocityField} also shows the location where the dust emission becomes optically thick along the midplane (that is, along the north-south direction) in our model. We adopt a parametrized dust opacity with a value of 3.5 cm$^2$ g$^{-1}$ at 850 $\mu$m and a spectral index $\beta$ of 0.25 \citep{Tobin2013}. At the frequency of the \hcop $J=4-3$ transition (356.734288 GHz; as observed), the dust becomes optically thick just outside the snowline. Even for the lowest \hcop transition ($J=1-0$ at 89.188523 GHz) the snowline coincides or falls just inside of the $\tau=1$ surface and will be hidden by the dust. 

\begin{figure*}
\centering
\includegraphics[width=\textwidth,trim={.5cm 13cm .5cm 2cm},clip]{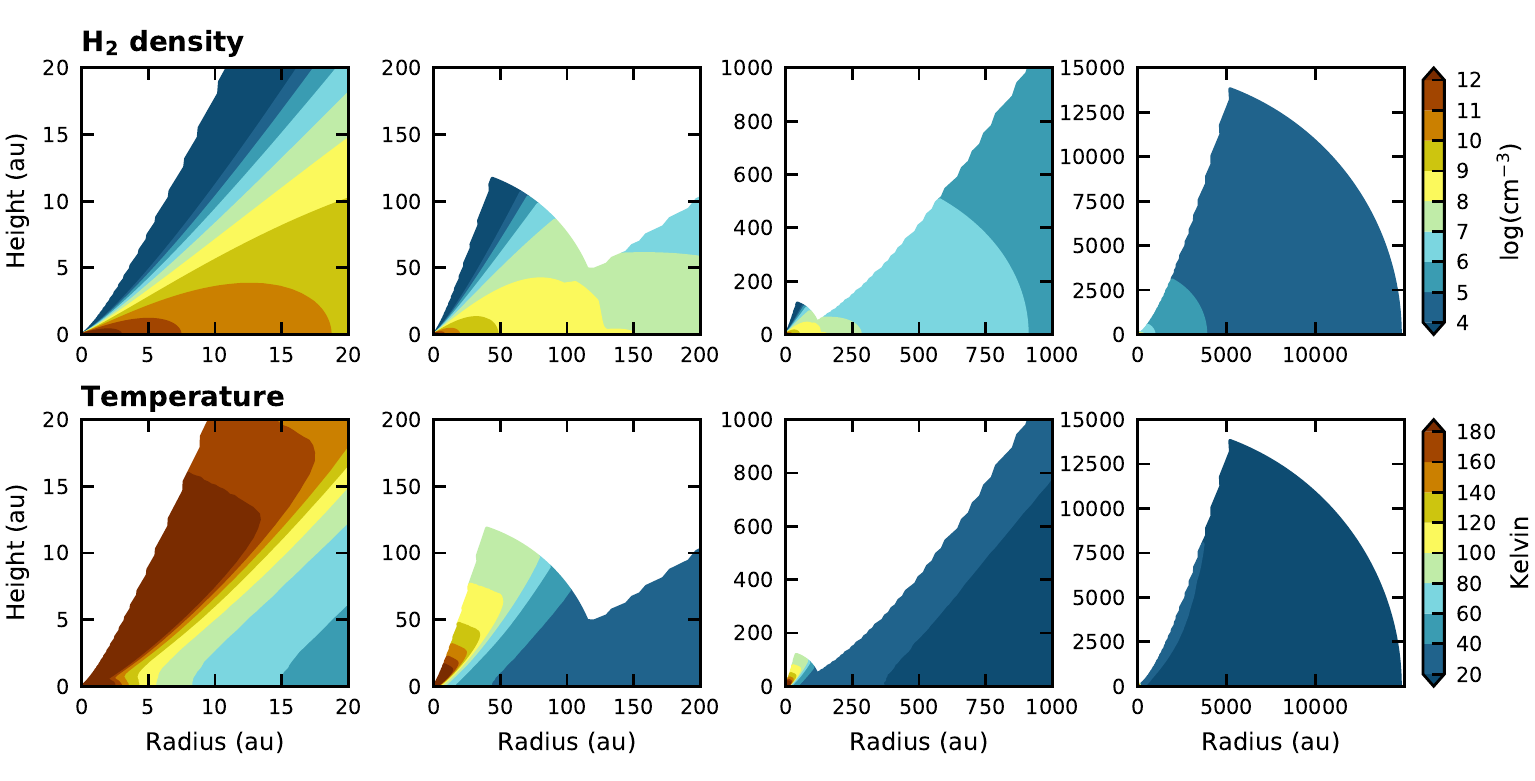}
\caption{Density (top panels) and temperture (bottom panels) structure for the disk and envelope of L1527 from the best fit model by \citet{Tobin2013}. From left to right, panels display larger spatial scales. The disk outer radius is 125 au. }
\label{fig:PhysicalStructure}
\end{figure*}

The temperature and density structure in the adopted physical model for L1527 is shown in Fig.~\ref{fig:PhysicalStructure}. This model was derived by \citet{Tobin2013} using a large grid of 3D radiative transfer models to model the thermal dust emission in the (sub-)millimeter, the scattered light $L^{\prime}$ image, and the multi-wavelength SED. The model includes a rotating infalling envelope \citep{Ulrich1976,Cassen1981} and a flared disk \citep[e.g.,][]{Hartmann1998}. In addition to a protostellar luminosity of 1.0 $L_\odot$, a luminosity of 1.75 $L_\odot$ is used to account for the accretion from the disk onto the protostar.

\section{Parametrized modeling of H$^{13}$CO$^+$ emission}\label{ap:ParametrizedModel}

\begin{figure*}
\centering
\includegraphics[trim={0cm 12.8cm 0cm 0cm},clip]{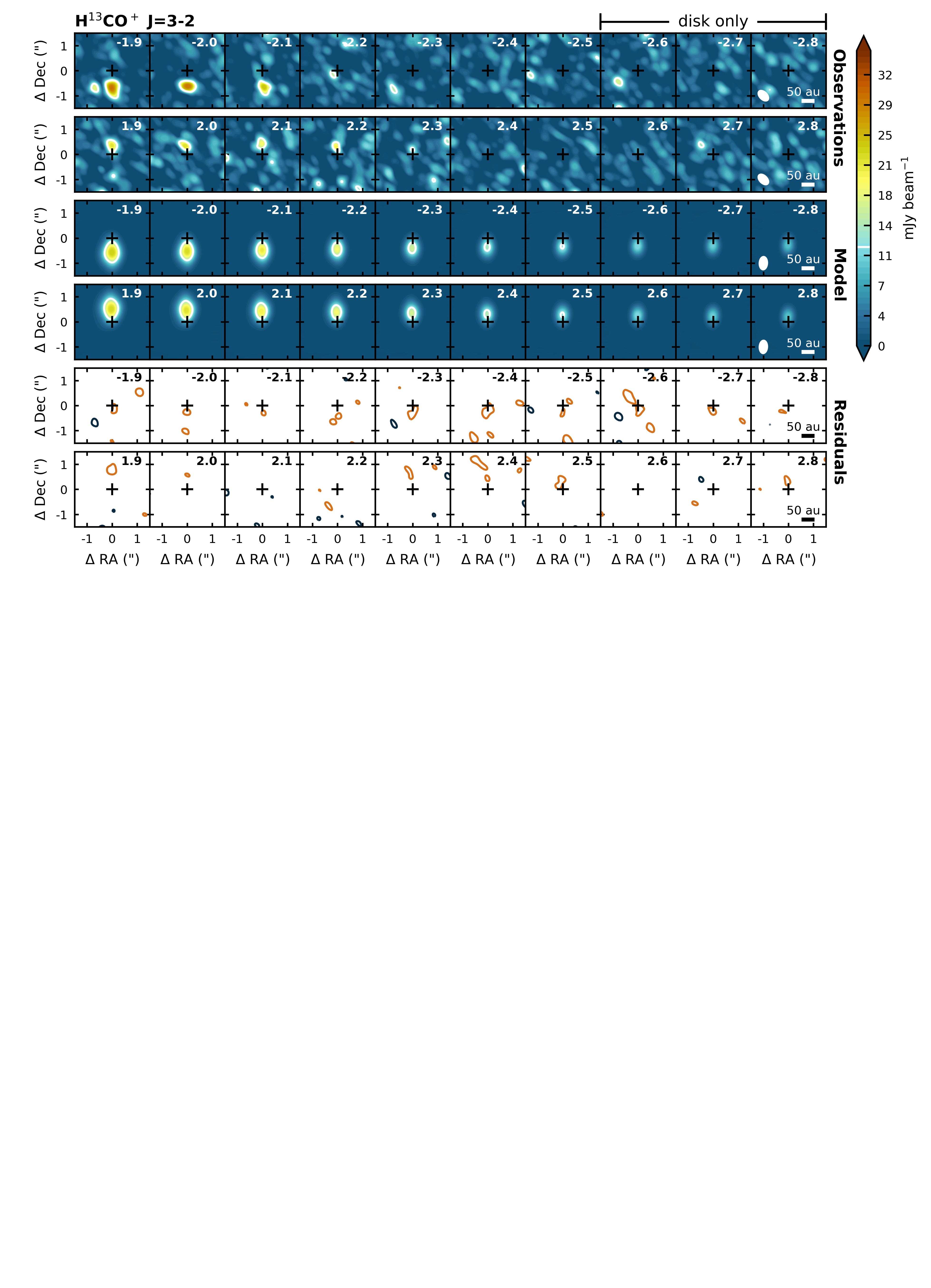}
\caption{As Fig.~\ref{fig:L1527_HCO+model}, but for \htcop emission. The model has a constant abundance of $3\times10^{-12}$. The depicted velocity range is slightly different from that shown for \hcop in Fig.~\ref{fig:L1527_HCO+model}, because the \htcop emission is only detected at slightly lower velocities. }
\label{fig:L1527_H13CO+model}
\end{figure*}

The \htcop observations can be reproduced by a constant abundance of $3\times10^{-12}$, as presented in Fig.~\ref{fig:L1527_H13CO+model}. This breaks the degeneracy between a \hcop model with an abundance of $2\times10^{-11}$ at radii $\leq$60 au and an abundance of $2\times10^{-10}$ at larger radii (as shown in Fig.~\ref{fig:L1527_HCO+model}) and a model with an abundance of  $2\times10^{-11}$ in the entire 125 au disk and an abundance of $1\times10^{-9}$ in the envelope, as the \htcop observations are in better agreement with the former \hcop abundance structure. 

\section{Chemical modeling of HCO$^+$ emission}\label{ap:Chemistry}

\begin{figure*}
\centering
\includegraphics[trim={0cm 0cm 0cm 0cm},clip]{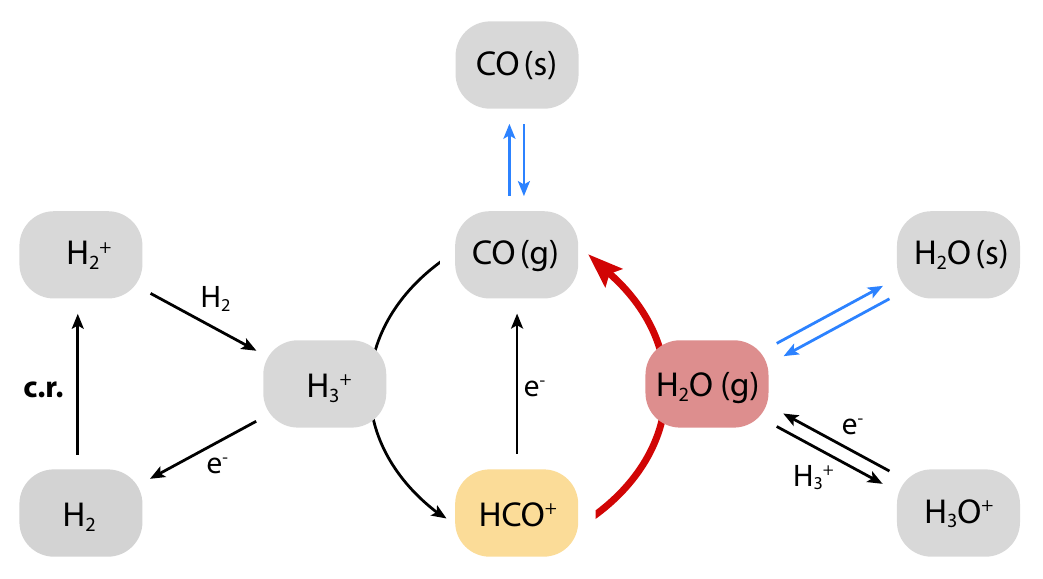}
\caption{Schematic representation of the chemical network used to model \hcop (orange). The reaction responsible for the anti-correlation between \hcop and H$_2$O is highlighted with a red arrow. Freeze out and thermal desorption reactions are marked by blue arrows, where `g' denotes gas-phase species and `s' denotes ice species. Ionization of H$_2$ by cosmic rays (c.r.) initiates the ion-neutral chemistry. \citet{Leemker2021} did not include the freeze out of CO, because their disks were too warm for CO ice. Parameters that are adjusted in this work are the initial CO and H$_2$O abundances and the cosmic ray ionization rate. }
\label{fig:ChemicalNetwork}
\end{figure*}

\begin{figure*}
\centering
\includegraphics[width=\textwidth,trim={.5cm 16.8cm .3cm 2.3cm},clip]{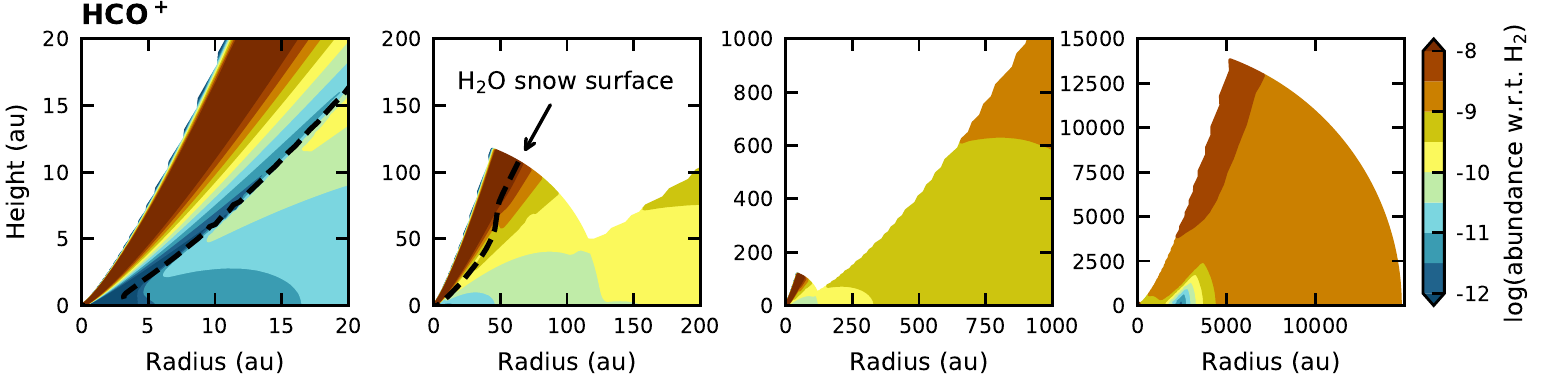}
\caption{Abundance structure for \hcop for the fiducial model (initial CO and H$_2$O abundances of 10$^{-4}$ and 10$^{-6}$, respectively, and a cosmic ray ionization rate of $10^{-18}$ s$^{-1}$) as shown in Fig.~\ref{fig:Chemistry_fiducial}, but after $10^6$ yr instead of $10^5$ yr. From left to right, panels display larger spatial scales. The disk outer radius is 125 au. The dashed line in the two left most columns marks the H$_2$O snow surface and the midplane snowline at 3.4 au.}
\label{fig:Chemistry_1e6yr}
\end{figure*}

\begin{figure*}
\centering
\includegraphics[width=\textwidth,trim={.5cm 16.8cm .3cm 2.3cm},clip]{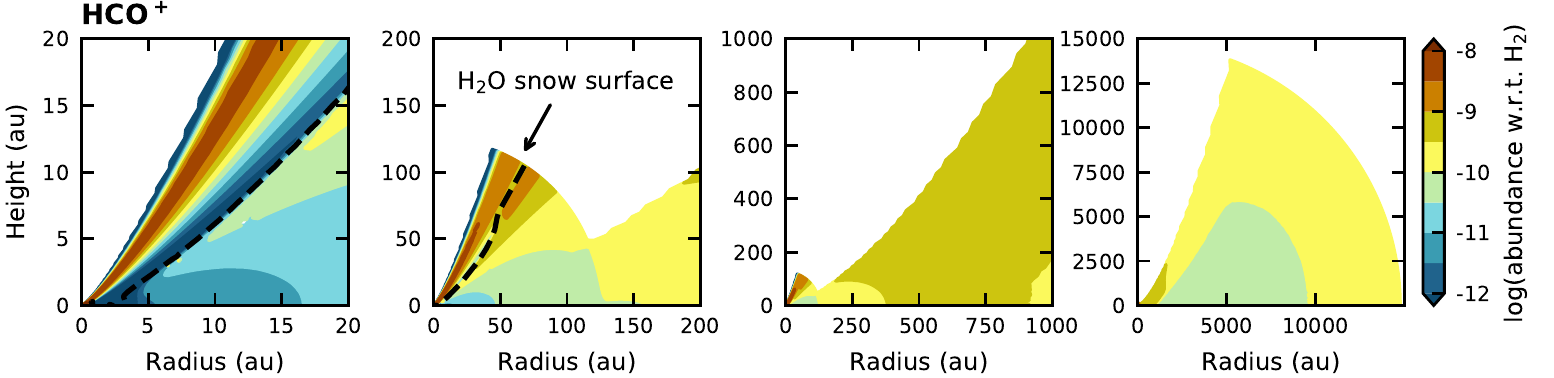}
\caption{Abundance structure for \hcop as shown in Fig.~\ref{fig:Chemistry_fiducial}, but for model with a lower initial CO abundance of 10$^{-6}$.}
\label{fig:Chemistry_lowCO}
\end{figure*}

\begin{figure*}
\centering
\includegraphics[width=\textwidth,trim={.5cm 16.8cm .3cm 2.3cm},clip]{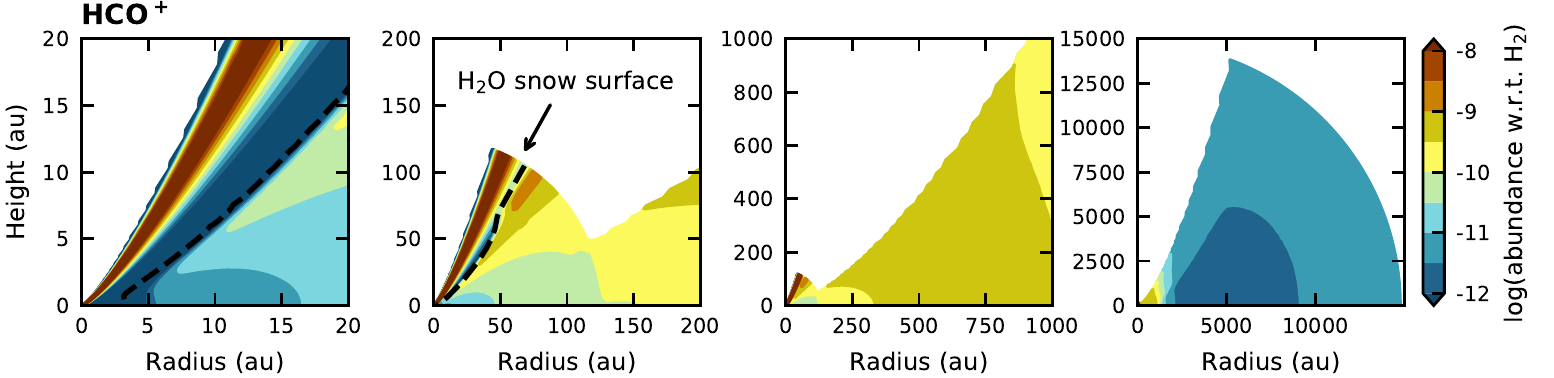}
\caption{Abundance structure for \hcop as shown in Fig.~\ref{fig:Chemistry_fiducial}, but for model with a higher initial H$_2$O abundance of 10$^{-4}$.}
\label{fig:Chemistry_highH2O}
\end{figure*}

\begin{figure*}
\centering
\includegraphics[width=\textwidth,trim={.5cm 16.8cm .3cm 2.3cm},clip]{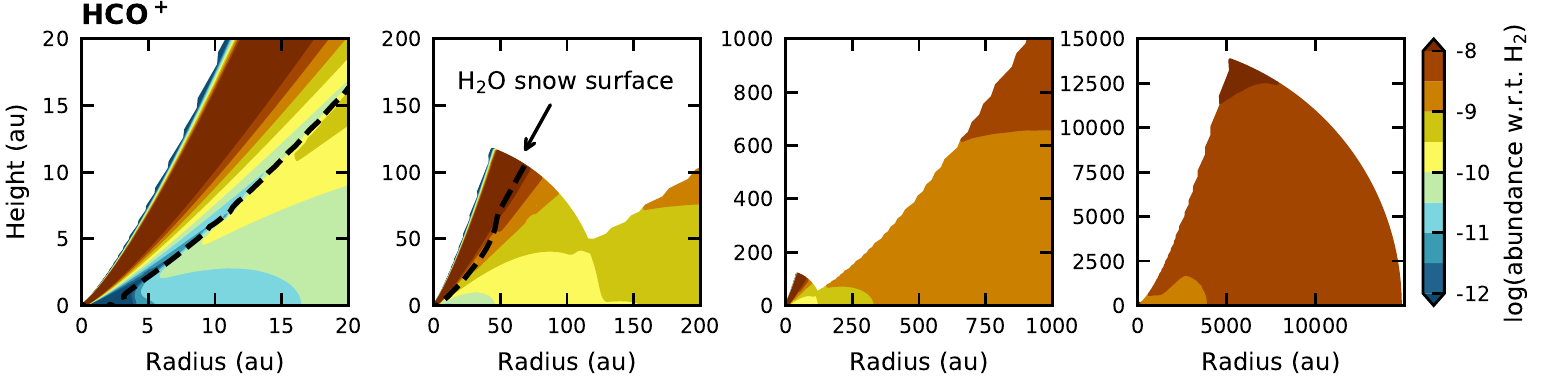}
\caption{Abundance structure for \hcop as shown in Fig.~\ref{fig:Chemistry_fiducial}, but for model with a canonical cosmic ray ionization rate of $10^{-17}$ s$^{-1}$.}
\label{fig:Chemistry_CR}
\end{figure*}

Figure~\ref{fig:ChemicalNetwork} shows a schematic of the chemical network used to model \hcop and its relation with the water snowline using the physical structure of L1527, as shown in Fig.~\ref{fig:PhysicalStructure}. The chemistry is evolved for a duration of $10^5$ yr, but for comparison, Fig.~\ref{fig:Chemistry_1e6yr} shows the \hcop abundance for the fiducial initial conditions of X(CO) = $10^{-4}$, X(H$_2$O = $10^{-6}$ and a CR rate of $10^{18}$ s$^{-1}$ after $10^6$ yr. The \hcop abundance after $10^5$ yr but for different initial conditions are presented in Figs.~\ref{fig:Chemistry_lowCO}-\ref{fig:Chemistry_CR}: a low CO abundance of $10^{-6}$ (Fig.~\ref{fig:Chemistry_lowCO}), a high H$_2$O abundance of $10^{-4}$ (Fig.~\ref{fig:Chemistry_highH2O}) and a CR rate of $10^{17}$ s$^{-1}$ (Fig.~\ref{fig:Chemistry_CR}).

\vspace{1cm}

\end{appendix}

\end{document}